\documentclass[aps,prb,reprint,superscriptaddress,floatfix]{revtex4-2}
\usepackage{amsmath} 
\usepackage{amssymb} 
\usepackage{physics} 
\usepackage{graphicx} 
\graphicspath{{figures/}} 
\usepackage{hyperref} 
\usepackage[version=4]{mhchem} 
\usepackage{bm} 
\usepackage{soul,color} 

\makeatletter
\newcommand{\colorcaption}[2][]{%
  \begingroup%
  \renewcommand{\@caption@fignum@sep}{ (Color online). }%
  \caption[#1]{#2}%
  \endgroup%
}
\makeatother

\DeclareGraphicsExtensions{.png,.pdf}

\begin{document}

\title{Mapping Spin Interactions from Conductance Peak Splitting in Coulomb Blockade}

\author{Eric D. Switzer}
\affiliation{Department of Physics, University of Central Florida, Orlando, Florida 32816, USA}
\author{Xiao-Guang Zhang}
\affiliation{Department of Physics, Center for Molecular Magnetic Quantum Materials and Quantum Theory Project, University of Florida, Gainesville, Florida 32611, USA}
\author{Volodymyr Turkowski}
\affiliation{Department of Physics, University of Central Florida, Orlando, Florida 32816, USA}
\author{Talat S. Rahman}
\email[Corresponding author email: ]{talat.rahman@ucf.edu}
\affiliation{Department of Physics, University of Central Florida, Orlando, Florida 32816, USA}

\date{\today}

\begin{abstract}
We investigate the transport properties of a quantum dot coupled to leads interacting with a multi-spin system using the generalized master equation within the Coulomb blockade regime.
We find that if two states for each scattering region electron manifold are included, several signatures of the interacting spin system appear in steady-state transport properties. 
We provide a theoretical mapping of differential conductance peak signatures and all spin Hamiltonian parameters related to the inclusion of excited state transitions between uncharged and charged electron manifolds.
Our predictions describe a scheme of only using a quantum dot and differential conductance to measure magnetic anisotropy, inter-spin exchange coupling, exchange coupling between the spin system and itinerant electron, and applied magnetic field response.
\end{abstract}

\maketitle

\section{Introduction}

    Determining the internal structure of a spin complex is important in both quantum information science (QIS) and spintronics. 
    Examples of useful spin complexes include molecular magnets (MMs) \cite{wilson06,wilson07,hill10,roch11,burzuri15}, coupled quantum dots (QDs) \cite{noiri18,nakajima19,noiri22}, and many-electron QDs \cite{leon21} because they possess properties like magnetic hysteresis, long spin-relaxation times, and protection against spin decoherence. 
    A complete description of their eigenspectrum involves mapping the properties of their internal structure onto effective spin model Hamiltonians. 
    Four common parameterized spin Hamiltonian terms for this purpose are magnetic anisotropy, exchange coupling between spin centers, exchange coupling of the spin centers with transitory electrons when the spin centers are placed between biased leads, and response to an applied magnetic field.
    
    There is significant work, utilizing a combination of theory and experiment, to match model spin Hamiltonian terms to experimentally-accessible transport measurements.
    Several commonly used techniques to characterize magnetic systems are electron paramagnetic resonance spectra measurements for crystalline MM complexes \cite{morenopineda14,hill03,gatteschi06}, magnetic susceptibility measurements \cite{motoyama96}, neutron inelastic scattering \cite{carretta03}, and magnetic circular dichroism spectroscopy \cite{piligkos09}.
    Characterizing systems important to QIS and spintronics involves the use of differential conductance measurements that exploit the Coulomb blockade (CB). Examples include probing exchange coupling and magnetic anisotropy for MM transistors such as $\text{N}@\text{C}_{60}$ \cite{roch11} and $\text{Fe}_{4}$ \cite{burzuri15}, the exchange coupling of two or more coupled QDs  \cite{waugh95,golden96,waugh96,you99,zhang12}, and the detection and manipulation of spin states \cite{noiri18,nakajima19,noiri22} for QD qubits.
    In the blockade, the flow of electrons is blocked by their Coulomb repulsion at low temperature and small bias voltage applied across leads connected through a central region \cite{averin86}.
    By constraining the dynamics to a single electron interacting with a complicated spin system, one can extract parameters based on repeated transport measurements.    

    Model approximations are often used for MMs, such as ignoring internal exchange coupling between spin centers and assuming a single spin $S$ (giant spin approximation). 
    For a certain class of spin complexes, this enables a tractable measurement scheme of some of the spin Hamiltonian terms \cite{roch11,burzuri15}.
    Other spin complexes, however, may not be described accurately by those approximations, such as $\text{Ni}_{4}$ single MMs (SMM) \cite{wilson06,wilson07} and $\text{Mn}_{3}$ dimer complexes \cite{hill10}.
    In some of the molecular cases, and in general with qubit read/write operations for tripartite spin systems \cite{switzer21,switzer22}, one must characterize all exchange couplings that are energetically relevant.
    Some approximations, such as ignoring a particular Hamiltonian term in MMs (e.g., between exchange coupling or magnetic anisotropy) cannot be made because they are both defined by the overlap of atomic orbitals belonging to the spin centers. 
    Changing one of the aforementioned parameters inevitably means that the other parameter also changes.
    
    Accurate measurements of all four parameters is then necessary in those cases to help screen materials for quantum architectures.
    In this paper we propose a scheme to map the four parameters of a particular class of spin complexes, namely exchange coupled spin dimers possessing magnetic anisotropy and coupled to an ``indirect measurement'' QD, using differential conductance and three experimentally-controlled parameters: anisotropically-applied magnetic field, bias voltage, and gate voltage.
    We rely on a rate equation-based theoretical approach of an electron transiting through a QD in the CB, as rate equations have been successful at identifying conductance peak features in transport spectra for systems consisting of an SMM placed between leads \cite{elste05,elste06,timm06,misiorny07apr,misiorny07aug,hymas19}.
    By including all four parameters in our spin model, we find that one can use the number and location of the peaks in differential conductance to determine each of the model's Hamiltonian parameters. 

    The paper is organized as follows. 
    In Sec.~\ref{sec:model}, we describe our model, write down the Hamiltonian and solve the generalized master equation to obtain closed equations for electronic current. 
    In Sec.~\ref{sec:results} we describe the role of the Hamiltonian terms in predicted differential conductance peaks. 
    Last, in Sec.~\ref{sec:discussion} and Sec.~\ref{sec:summary}, we summarize the results and discuss the experimental scheme to measure the parameters for each Hamiltonian term.

\section{Model and Hamiltonian}
\label{sec:model}
\begin{figure}[!ht]
        \includegraphics[width=\columnwidth]{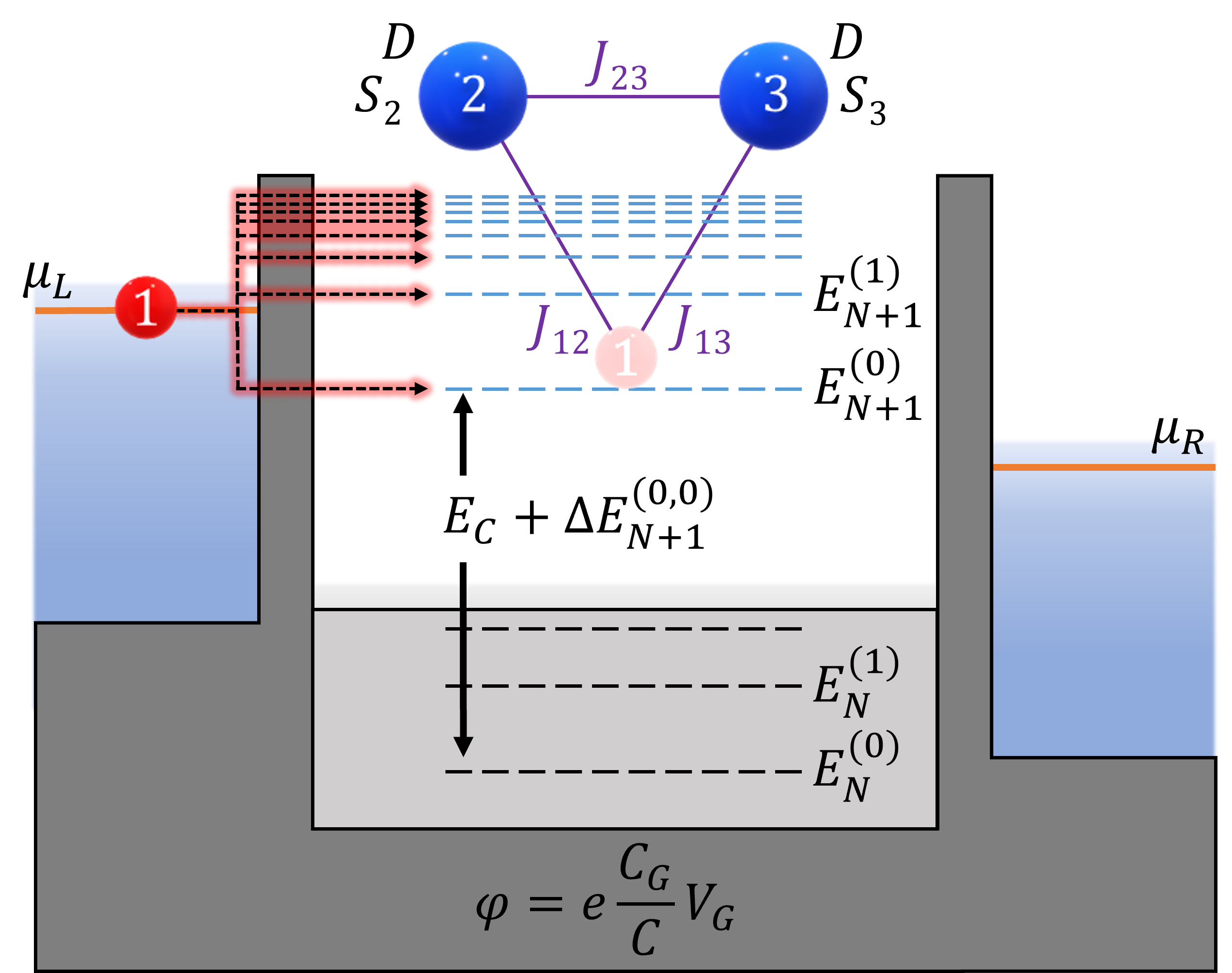}
        \colorcaption{\label{fig:schematic}
        Schematic of the system consisting of a central scattering region, containing an $S_{2,3}=1$ spin dimer complex interacting through an exchange interaction $J_{23}$, coupled to polarized leads at temperature $T$. 
        The central region’s eigenenergy levels are tuned via gate voltage $V_{G}$ so that the charged ground state energy $E_{N+1}^{(0)}$ with $N+1$ electrons is aligned with the unbiased leads, i.e., $E_{N+1}^{(0)}=\mu_{L}=\mu_{R}=0$ eV. 
        Applying a symmetric bias voltage as shown enables transport of a single electron (particle 1) through the $N + 1$ electron states. 
        Once the electron has transported into the central region, and prior to leaving the central region, additional exchange interactions $J_{1i}$ couples the electron’s spin to the dimer.
        }
    \end{figure}

    A three-terminal setup consisting of a source and drain electrode, and a gate, is found in many nanosized devices, including three-terminal coupled QDs \cite{noiri18,nakajima19,noiri22}, nuclear and molecular spin qubit transistors \cite{thiele14,najafi19,biard21}, and magnetic molecule tunneling junctions \cite{gonzalez07,gonzalez08,hymas20thesis,zhang21}. 
    We consider a hybrid of the aforementioned setups by modeling a central region consisting of a QD influenced by two spin particles, connected to three terminals.
    This model is functionally equivalent to the models explored in \cite{kim22pb,kim22pla}, but instead of exploring timescales in which coherence can be tracked, we focus on timescales in which incoherent transport is measured.
    The source and drain electrodes enable transport of an itinerant electron into the QD where it interacts with other spins via a spin exchange interaction.   
    The energy levels of the central region are adjusted by the third terminal to bring the system into the CB regime.
    The overall model is shown schematically in Fig.~\ref{fig:schematic}.

    \subsection{Hamiltonian}
        The total Hamiltonian is,
        \begin{align}
        \label{eqn:hamiltonian_total}
            \mathcal{H}=\mathcal{H}_{\text{leads}}+\mathcal{H}_{\text{leads}-\text{C}}+\mathcal{H}_{\text{C}},
        \end{align}
        where each term in Eq.~(\ref{eqn:hamiltonian_total}) is explained as follows.
        The energy of the electrons on the leads $\mathcal{H}_{\text{leads}}$ is,
        \begin{align}
        \label{eqn:hamiltonian-leads}
            \mathcal{H}_{\text{leads}}=\sum_{\alpha k\sigma}\left(\epsilon_{\alpha k}+\mu_{\alpha}\right)\hat{c}^{\dagger}_{\alpha k\sigma}\hat{c}_{ \alpha k\sigma},
        \end{align}
        where $\hat{c}^{\dagger}_{\alpha k\sigma}$ creates an electron in the left and right lead, $\alpha=L,R$, respectively, with energy $\epsilon_{\alpha k}$ relative to the chemical potential of the lead $\mu_{\alpha}$, momentum $k\equiv\vb{k}$, and spin state $\sigma$ projected on the $z$-axis set by the easy-axis of the zero-field splitting term described later. 
        We set the zero of the lead's chemical potential to the ground-state of the $N+1$ electron manifold of the central region, and the bias voltage is applied symmetrically such that $\mu_{L}=V_{b}/2$ and $\mu_{R}=-V_{b}/2$.
        The coupling of the leads to the central region can be described by the hybridization term $\mathcal{H}_{\text{leads}-\text{C}}$,
        \begin{align}
            \label{eqn:hamiltonian-lead-center-IQD}
            \mathcal{H}_{\text{leads}-\text{C}}=\sum_{\alpha k\sigma n}\left(t_{\alpha k \sigma}\hat{c}^{\dagger}_{ \alpha k\sigma}\hat{d}_{n\sigma}+h.c.\right),
        \end{align}
        where $t_{\alpha k \sigma }$ is the tunneling amplitude of an electron with momentum $k$ at lead $\alpha$ to level $n$, $\hat{d}^{\dagger}_{n\sigma}$ creates an electron with spin $\sigma$ on the central region's eigenstate $n$, and we have assumed that the hopping rate is independent of $n$.
        
        The central region Hamiltonian has the form,
        \begin{align}
            \mathcal{H}_{\text{C}}&=\mathcal{H}_{eS}+\mathcal{H}_{23}+\mathcal{H}_{A}+\mathcal{H}_{Z}+\mathcal{H}_{E}+\mathcal{H}_{G}.
        \end{align}
        $\mathcal{H}_{eS}$ refers to the exchange interaction of the electron in the QD with the two spin particles in the central region,
        \begin{align}
            \label{eqn:dimer-kondolike}
            \mathcal{H}_{eS}&=\delta_{N_{e},1}\frac{1}{2}J_{1i}\sum_{in\mu\mu^{\prime}}\hat{\vb{S}}_{i}\cdot\hat{d}^{\dagger}_{n\mu}\hat{\bm{\sigma}}_{\mu\mu^{\prime}}\hat{d}_{n\mu^{\prime}},
        \end{align}
        where $N_{e}$ is the number of extra electrons in the central region (i.e., the charge state), $J_{1i}$ is the exchange interaction between an electron and the spin particles (assumed to be the same strength for each spin particle $i$), $\vb{S}_{i}$ is the spin operator for spin particle $i$, and $\hat{\bm{\sigma}}_{\mu\mu^{\prime}}$ is the corresponding $\mu,\mu^{\prime}$ matrix element of the $s=\frac{1}{2}$ Pauli matrix.
        This form of the exchange interaction is derived by extending the single impurity Anderson model \cite{anderson66} to a two-impurity Anderson model, and transforming into the low energy regime by means of the Schrieffer-Wolff transformation \cite{schrieffer66}.
        The $\mathcal{H}_{23}$ term similarly refers to interactions between the two spin centers,
        \begin{align}
            \label{eqn:23interaction}
            \mathcal{H}_{23} &= J_{23}\hat{\vb{S}}_{2}\cdot\hat{\vb{S}}_{3},
        \end{align}
        where $J_{23}$ characterize the exchange interaction between spin particle 2 and 3.
        
        Next, the $\mathcal{H}_{\text{A}}$ term originates from the spin-orbit interaction of one or more unpaired electrons in spin $S>\tfrac{1}{2}$ impurities, and describes the effective zero-field splitting (and magnetic anisotropy), of spin particle 2 and 3,
        \begin{align}
            \label{eqn:anisotropy}
            \mathcal{H}_{A}&=D\left(\hat{S}^{z}_{2}\hat{S}^{z}_{2}+\hat{S}^{z}_{3}\hat{S}^{z}_{3}\right).
        \end{align}
        Here $D$ is the magnitude of the uniaxial anisotropy strength, with the $z$ direction determined by the preferential direction of the easy axis, and in general is dependent on the charge state.
        The $\mathcal{H}_{Z}$ term represents the applied magnetic field,
        \begin{align}
            \label{eqn:zeeman}
            \mathcal{H}_{Z}&=\mu_{B}\left(\delta_{N_{e},1}g_{1}\hat{\bm{\sigma}}+g_{2}\hat{\vb{S}}_{2}+g_{3}\hat{\vb{S}}_{3}\right)\cdot\vb{B},
        \end{align}
        where $g_{i}$ is the isotropic $g$ factor for spin particle $i$, $\mu_{B}$ is the Bohr magneton, and $\vb{B}=B_{x}\hat{x}+B_{y}\hat{y}+B_{z}\hat{z}$ is an applied magnetic field.
        As indicated in this Hamiltonian term, the magnetic field is locally applied within the central region, and applies to the electron only if it transits the central region.
        
        The last two Hamiltonian terms are electrostatic in nature and describe the charging energy of the central region and the applied gate voltage \cite{kouwenhoven97},
        \begin{align}
            \label{eqn:charge-gate}
            \mathcal{H}_{E}+\mathcal{H}_{G}&=\frac{N^{2}_{e}}{2}E_{C}-N_{e}e\frac{C_{G}}{C}V_{G},
        \end{align}
        where $E_{C}$ is the charging energy $e^{2}/2C$, $e$ is the charge of the electron, $C$ is the total capacitance across the barriers, $C_{G}$ is the capacitance of the central region connected to the gate, and $V_{G}$ is the gate voltage. 
        In this work, we do not include $N+2$ and $N-1$ electron manifolds (i.e., $N_{e}=2$ and $N_{e}=-1$, respectively) because those manifolds are assumed to be energetically unfavorable. 
        We also rescale $V_{G}$ to highlight the net effect of the gate voltage on transport properties by performing the transformation $\frac{C_{G}}{C}V_{G}\rightarrow V_{G}$. 
        Under the manifold assumption, the electrostatic terms become,
        \begin{align}
            \label{eqn:charge-gate-simp}
            \mathcal{H}_{E}+\mathcal{H}_{G}&=\frac{1}{2}E_{C}-eV_{G},
        \end{align}
        for the $N+1$ electron manifold, and zero for the $N$ electron manifold.
        As will be useful later, we define $\Delta E^{(i,j)}_{N,N+1}\equiv E^{i}_{N}-E^{j}_{N+1}$ as the energy difference of the $i$'th and $j$'th eigenstate of the $N$ and $N+1$ manifold, where $i=0$ defines the ground state of that manifold. 
        The magnetic parameters chosen for our model are on the order of $\text{cm}^{-1}$, as is common with MM systems. Similarly for parameters that can be easily tuned experimentally, the fields are on the order of $T$, while the electrostatics are on the order of mV to access useful transport properties.

    \subsection{Model Details} 
       We work in the regime appropriate for single-charge dynamics, i.e., the CB regime, along with weak lead-central region interactions.
       As a result, third-order and higher terms of the lead-central region perturbation will not contribute much to the dynamics of the system's density matrix. 
       This allows consideration up to second-order in the perturbation, enabling access to a tractable solution of electronic current.  
       
       The system's density matrix can be decoupled into two parts,
        \begin{align}
            \label{eqn:dm-bornapprox}
            \rho_{I}(t) = \rho_{C}(t)\rho_{\text{leads}}(0),
        \end{align}
        where $\rho_{I}(t)$ is the density matrix of the system in the interaction picture, $\rho_{C}(t)$ is the density matrix of the central region, and $\rho_{\text{leads}}(0)$ is the density matrix of the leads before the perturbation term is turned on.
        As a consequence of the weak lead-central region interaction and CB, the time-dependent spin entanglement predicted in Ref.~\cite{switzer21,switzer22} will not be accessible. 
        In order for equation Eq.~(\ref{eqn:dm-bornapprox}) to hold, the entanglement information between an electron coupled with the central region must be lost after some time $t=t_{c}$, where $t_{c}$ is on the order of the coherence time of the system. 
        
        Because the reservoir is split into two leads, they are presumed to not interact with each other nor possess spin levels that interact with each other. 
        The density matrix of the leads can then be separated by lead and by spin,
        \begin{align}
            \label{eqn:dm-sepleads}
            \rho_{\text{leads}}(0)
            &= \rho_{L\uparrow}(0) \otimes \rho_{L\downarrow}(0) \otimes \rho_{R\uparrow}(0) \otimes \rho_{R\downarrow}(0).
        \end{align}
        The constant density matrix of the leads essentially means that the central region does not have an appreciable effect on the leads, and the leads maintain a thermal equilibrium. 
        This is a statement of irreversibility of the system considered in this work.
        
        Next we assume the Markov approximation, in that the behavior of the central region is not related to its behavior at any past time. 
        This is justified because we assume that the coupling of the central region to the leads is at least strong enough to dampen any long-term correlations. 
        To allow the Markov approximation to hold, we consider times longer than the natural frequency of oscillations between two central region energy states $n$ and $n'$, $t \gg \hbar / \left|\omega_{n'n}\right|$.
        
        We assume that the coupling of the central region to the environment of the leads is weak enough in which the change of the total density matrix in the interaction picture is slow. 
        By choosing a long time in which the Redfield relaxation tensor is approximately independent of time, we use the secular approximation by maximizing the exponential factor in front of the Redfield relaxation tensor to be unity. 
        The surviving secular terms are bound by energies that satisfy $\omega_{n'n}-\omega_{N'N}=0$ where the difference is defined between the natural frequency of between two states of the central region $\omega_{n'n}$ and the natural frequency of two states in the leads $\omega_{N'N}$.

    \subsection{Generalized Master Equation}
        We next follow the well-known Fermi golden rule approach to Coulomb blockade transport and construct the generalized master equation. 
        We assume that quasiparticle lifetime $\tau_{q}$ within those manifolds greatly depends on their relative energies, i.e., $\tau_{q} \ll \hbar/|\Delta E^{(i,j)}_{N,N+1}|$.
        By inspecting the magnitude of $\Delta E^{(0,2)}_{N,N+1}$, the quasiparticle lifetime of this excitation is likely too short to participate in transport across the leads. 
        Higher-order excitations with energy differences $\Delta E^{(i,j)}_{N,N+1}$ for $i > 1$ or $j > 1$ do not participate in electron transport if the transport channel involving $\Delta E^{(0,2)}_{N,N+1}$ does not participate.
            
        For our model, the generalized master equation is then,
        \begin{align}
            \nonumber
            \dot{\rho}_{n'n}(t) &= \frac{i}{\hbar}\left[\rho(t),\mathcal{H}_{0}\right]_{n'n}+\delta_{n'n}\sum_{m,n\neq m}\rho_{mm}(t)W_{n'm}\\
            \label{eqn:gme}
            &\;\;\;\;-\gamma_{n'n}\rho_{n'n}(t),
        \end{align}
        where $\rho(t)$ refers to the \emph{central region's} density matrix, $\rho(t) = \rho_{C}(t)$, and with the notation $\mathcal{H}_{n'n}\equiv\mel{n'}{\mathcal{H}_{0}}{n}$, $\rho_{n'n}\equiv\mel{n'}{\rho(t)}{n}$. 
        Each term in Eq.~(\ref{eqn:gme}) is explained as follows.

        The first term on the right-hand side of Eq.~(\ref{eqn:gme}) is the usual evolution of the central region's Hamiltonian containing the $QD$ and multi-spin system, and the lead Hamiltonian. 
        The dynamics of the system due to the coupling of the leads is given in the next two terms.
        The second term contains the transition rates between eigenstates of the system $W_{n'm}$ from state $\ket{m}$ to $\ket{n'}$ and is a sum of the contributions from each lead and spin polarization, i.e., $W_{n'm} = \sum_{\alpha\sigma}W^{\alpha\sigma}_{n'm}$. These rates are derived in Appendix \ref{appendix:density-matrix}. The result for the $N\rightarrow N+1$ (``absorption'') electron manifold transitions are,
        \begin{align}
            \label{eqn:trans-rate-absorb}
            W^{\alpha\sigma}_{c_{i}u_{j}} &= w_{\alpha\sigma}\nu_{\alpha\sigma}\abs{\mel{c_{i}}{\hat{c}^{\dagger}_{\alpha \sigma}}{u_{j}}}^{2}f_{\alpha}(\Delta E^{(i,j)}_{N+1,N}),
        \end{align}
        where $w_{\alpha\sigma}=2\pi\abs{t_{\alpha\sigma}}^{2}D(E_{f})/\hbar$ are the lead and polarization dependent transition rate constants, $f_{\alpha}(E)$ is the Fermi function of lead $\alpha$, $D(E_{f})$ is the density of states at the Fermi energy, and $\nu_{\alpha\sigma}$ is the fractional polarization of lead $\alpha$ constrained to the normalization condition $\nu_{\alpha\uparrow}+\nu_{\alpha\downarrow}=1$. 
        For example, the leads can be chosen to be fully polarized, e.g,  $\nu_{L\downarrow} = \nu_{R\uparrow} = 1.0$ and $\nu_{R\downarrow} = \nu_{L\uparrow} = 0.0$, or non-polarized, i.e., $\nu_{\alpha\sigma} = 0.5 \;\forall \;\alpha, \sigma$.
        The transition rates for $N+1\rightarrow N$ (``emission'') electron manifold transitions are similarly,
        \begin{align}
            \label{eqn:trans-rate-emit}
            W^{\alpha\sigma}_{u_{i}c_{j}} &= w_{\alpha\sigma}\nu_{\alpha\sigma}\abs{\mel{u_{i}}{\hat{c}_{\alpha \sigma}}{c_{j}}}^{2}\left(1-f_{\alpha}(\Delta E^{(j,i)}_{N+1,N})\right).
        \end{align}
        
        The last term on the right-hand side of Eq.~(\ref{eqn:gme}) contains a damping factor $\gamma_{n'n}$, also derived in Appendix \ref{appendix:density-matrix}. This factor is a consequence of the lead's interaction with the central region, and is defined for states $n' \ne n$ as,
        \begin{align}
            \label{eqn:gamma-with-T2}
            \gamma_{n'n}=\frac{1}{2}\sum_{m}\left(W_{mn'}+W_{mn}\right)+\frac{1}{T_{2}},
        \end{align}
        where $T_{2}$ is the spin decoherence time. 
        This $T_{2}$ time can be due to a variety of sources such as spin-spin coupling with the system and the reservoir, e.g., between the magnetic moment of the spin particles and the magnetic moment of the atoms in the surrounding substrate.
        $T_{2}$ times have a range of magnitudes depending on the spin system of interest at low temperatures $T\approx 1$ K, such as $10^{-7}$ s for magnetic adatoms on surfaces, $10^{-7}$ to $10^{-5}$ s for QDs, and $10^{-4}$ to $10^{-1}$ s for systems of donor electrons embedded in silicon \cite{delgado17}.

        To produce relevant predictions from the generalized master equation, we look at a time range in which the overall relaxation time due to transitions $\tau$, e.g. phonon-induced, is much longer than the decay of the off diagonal elements $\tau_{d}=1/\gamma_{mm'}$. 
        This means that for $\rho_{n'n}(t)\propto e^{-t/\tau_{d}}\rightarrow\dot{\rho}_{n'n}(t)=-(1/\tau_{d})e^{-t/\tau_{d}}$, so we choose a long enough time such that $t >> \tau_{d}$ so that $\dot{\rho}_{n'n}(t)\rightarrow 0$. 
        We find that to in order to have non-zero electronic current, this condition is equivalent to the requirement that the off-diagonal terms of each electron manifold must be non-zero, agreeing with the conditions of non-zero current of a similar model in Ref.~\cite{gonzalez07,gonzalez08}. 
        Finally, the diagonal elements of the differential density matrix are solved by assuming the steady-state case, i.e. choosing some time $t_{s}>>1/W_{mm'}$ to obtain closed equations of the density matrix elements $\rho_{nn}\equiv \rho_{nn}(t_{s})$.
        We define the current through the central region as the transition from the charged to uncharged state across lead $\alpha$ and polarization $\sigma$. 
        The long-time steady-state current is then,
        \begin{align}
            \label{eqn:steady-state-current}
            I_{T} &= \left(I_{R\uparrow}-I_{L\uparrow}\right)+\left(I_{R\downarrow}-I_{L\downarrow}\right),
        \end{align}
        where,
        \begin{align}
            \nonumber
            I_{\alpha\sigma} &= e\left(W^{\alpha\sigma}_{u_{0}c_{0}}+W^{\alpha\sigma}_{u_{1}c_{0}}\right)\rho_{c_{0}c_{0}}\\
            \label{eqn:electric-current-bylead}
            &\;\;\;\;+e\left(W^{\alpha\sigma}_{u_{0}c_{1}}+W^{\alpha\sigma}_{u_{1}c_{1}}\right)\rho_{c_{1}c_{1}},
        \end{align}
        is the steady-state current through lead $\alpha$ with spin polarization $\sigma$.

\section{Results}
\label{sec:results}

    \subsection{Field-Dependent Energy Level Shifts}
    We first choose a system with parameters that will incorporate all dynamics presented in prior sections, while simplifying some parameter choices in order highlight the role of each interaction in the total Hamiltonian. 
    To this end, we assume an easy axis anisotropy for $S_{2}$ and $S_{3}$,  $D=-0.6\;\text{cm}^{-1}$, an isotropic antiferromagnetic coupling between the centers with $J_{23}=0.6\;\text{cm}^{-1}$, an isotropic ferromagnetic coupling of the itinerant electron with each center,  $J_{1i}=-0.8\;\text{cm}^{-1}$, and a charging energy of $E_{C}=1\;\text{meV}$. 
    To obtain non-zero current within a chosen bias window, we set the charge-state decoherence to $10\;\mu\text{eV}$.  
    The gate voltage is initially set to $V_{G}=0\;\text{mV}$.
    For simplicity, the $g$ factors of the three spin particles are assigned the same value $2.2$. 
    A small longitudinal magnetic field, parallel to the zero-field easy axis, is applied to aid in numerical convergence, $B_{z}=0.1$ mT. 
    The temperature is set to be sufficiently low for the CB to hold, at $T=0.1$ K. 
    For our figures, we choose non-polarized leads. 
    The tunneling rates are chosen to be $w_{\alpha\sigma}=10$ GHz.
    The spin decoherence time is set to be on the order of some magnetic molecules at $T_{2}=5.0$ ns \cite{delbarco2004}.
    
    We diagonalize each block of the uncharged and charged sectors for various choices of applied transverse magnetic field $B_{x}$ and $B_{y}$, $B_{x}$, $B_{y} \ge 0$. 
    We find that the energetics and transport behavior of this system are dependent on the magnitude of the applied transverse magnetic field, and not the direction of the field on the plane perpendicular to the shared easy axis of spin particle 2 and 3. 
    This contrasts the energy differences and asymmetric transport as a function of applied magnetic field predicted in Ref.~\cite{gonzalez07,gonzalez08} because our transport equations are derived to use the eigenstates of the central region Hamiltonian, allowing us to consider additional transitions. 
    We also do not see a dependence on the direction of the transverse field because there is no energetic preference towards a particular direction in any of the spin Hamiltonian terms (e.g., the lack of a a $E(\hat{S}^{2}_{x}-\hat{S}^{2}_{y})$ term).
    
\begin{figure}[!ht]
        \includegraphics[width=\columnwidth]{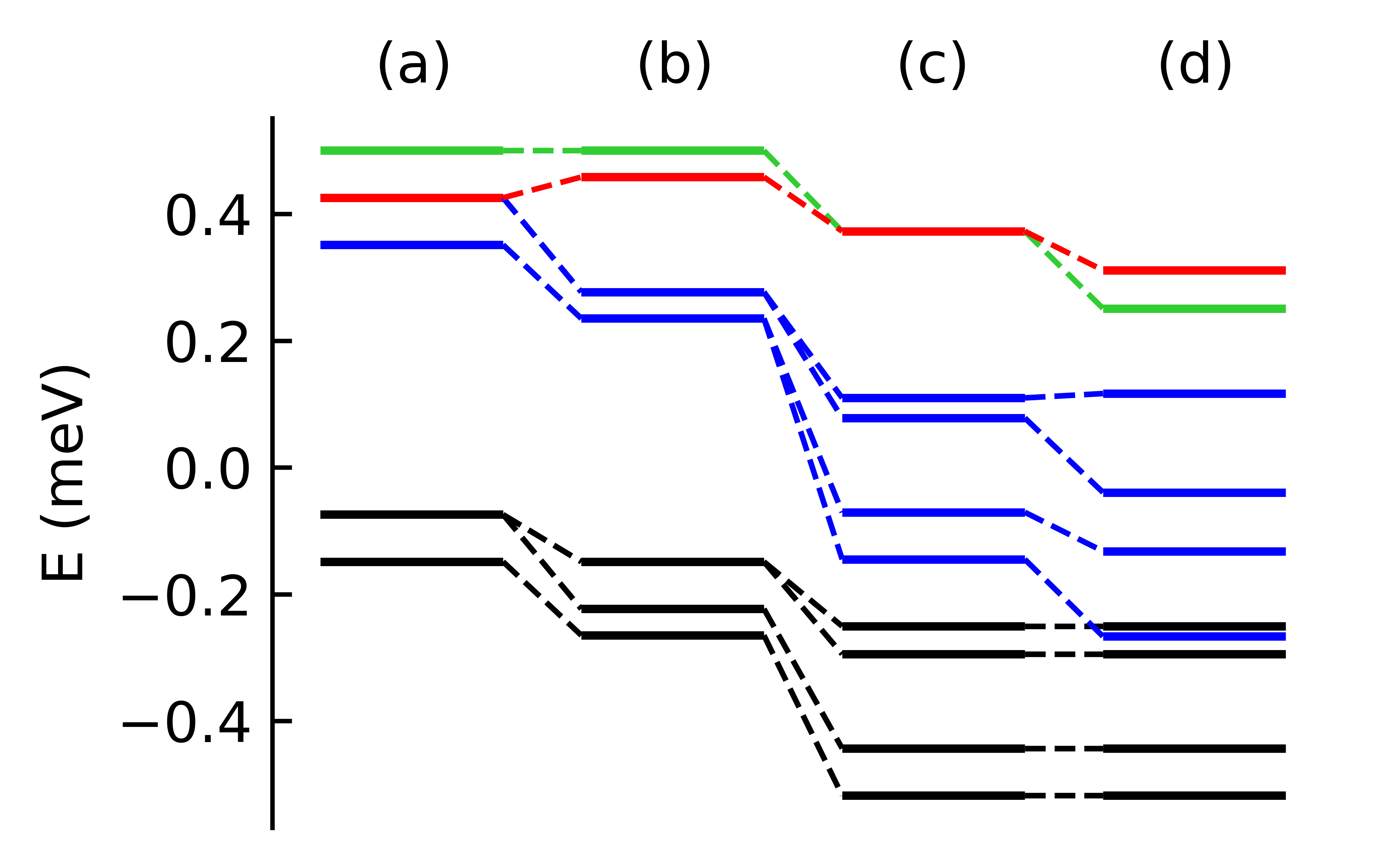}
        \colorcaption{\label{fig:explanation}
        First four energy levels of the $N$ (black) and $N+1$ (blue) central region electron manifolds. 
        Energy differences $\Delta E^{(0,0)}_{N+1,N}$ (green) and $\Delta E^{(1,1)}_{N+1,N}$ (red) are also plotted. 
        (a) Only $\mathcal{H}_{23}$ and $\mathcal{H}_{E}$ interactions are turned on, with $J_{23}=0.6\;\text{cm}^{-1}$ and $E_{C}=1\;\text{meV}$. 
        (b) The zero-field splitting term $\mathcal{H}_{A}$ is turned on with $D=-0.6\;\text{cm}^{-1}$. 
        (c) The applied magnetic field term $\mathcal{H}_{Z}$ is turned on with a sufficiently high field, $B_{x}=0.5\;\text{T}$, resulting in degenerate energy differences. 
        (d) Finally, the exchange interaction interaction $\mathcal{H}_{eS}$ is turned on, $J_{1i}=-0.8\;\text{cm}^{-1}$, breaking the degeneracy.
        }
    \end{figure}

    The transition rates of Eq.~(\ref{eqn:trans-rate-absorb}) and Eq.~(\ref{eqn:trans-rate-emit}) used in the transport equation crucially depend on the energy differences between electron manifolds $\pm\Delta E^{(i,j)}_{N,N+1}$. 
    The validity of the form of the transition rates is also dependent on maintaining the CB which in turn is dependent on the energy levels of the central region and applied bias. 
    Each contribution in the total Hamiltonian, then, will play a role in the transport equations. 
    To elaborate on the roles of the Hamiltonian terms, we inspect the first four energy levels for each charge sector in Fig.~\ref{fig:explanation} for a fixed transverse field of $B_{trans}=2.0$ T.
    For these choices of parameters, if the $\mathcal{H}_{23}$ and $\mathcal{H}_{E}$ interactions are turned on and the other interactions are off, the uncharged sector contains a non-degenerate ground state, and a three-fold degenerate first excited state.
    The ground and first-excited state of the charged sector are each two-fold degenerate. 
    When the zero-field splitting $\mathcal{H}_{A}$ is turned on, the three-fold degeneracy of the uncharged sector is broken. 
    The energy differences between ground and excited states of both manifolds are shifted as a result, and appear to approach a shared value. 
    Turning on the transverse magnetic field $H_{Z}$ completely breaks the degeneracy of both charge sectors. 
    The resulting energy differences, however, become degenerate. 
    We find that this degeneracy occurs around 0.2 T for the parameters used for Fig.~\ref{fig:explanation}, and persists for fields up to 2.0 T.
    When the exchange interaction of the electron is included, the energy difference symmetry is broken.
    
\begin{figure}[!ht]
        \includegraphics[width=\columnwidth]{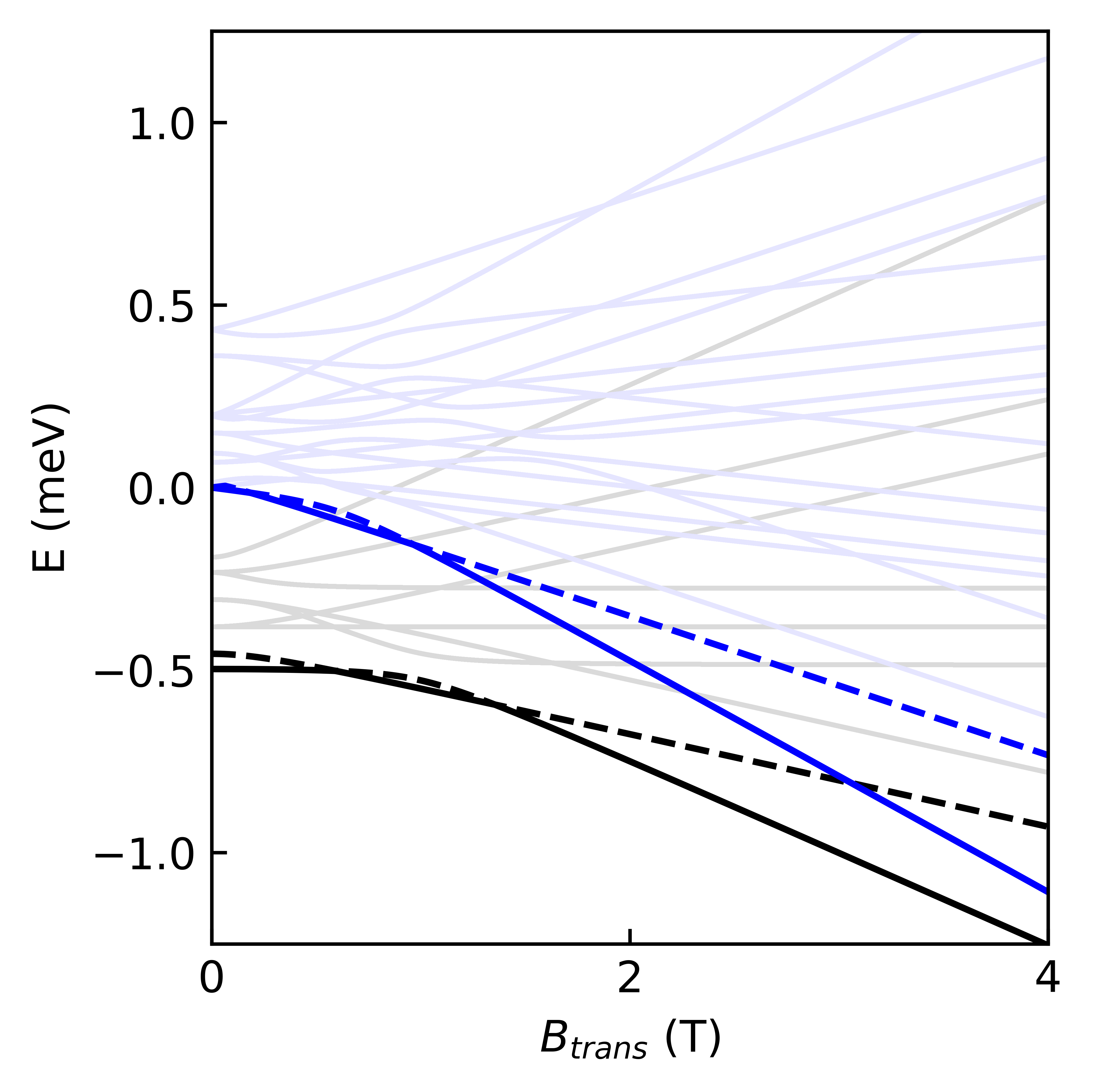}
        \colorcaption{\label{fig:energetics}
        Energy levels $E$ of the $N$ (black) and $N+1$ (blue) central region electron manifold. 
        The ground (solid), first excited (dashed), and higher-order (light solid) states are plotted using the parameters given in the text, as a function of applied transverse magnetic field $B_{trans}$.
        }
    \end{figure}

\begin{figure}[!ht]
        \includegraphics[width=\columnwidth]{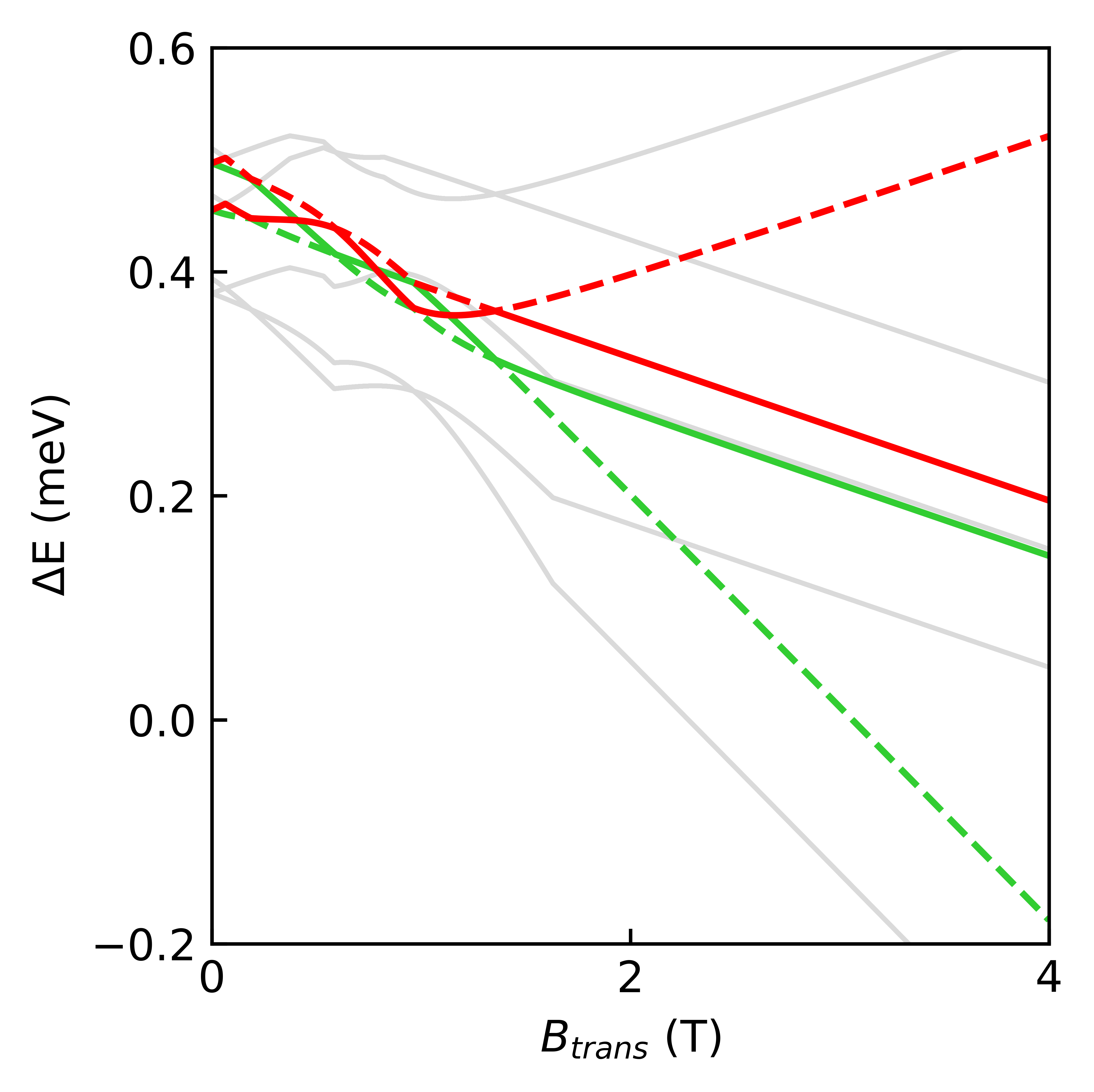}
        \colorcaption{\label{fig:splits}
        Energy difference $\Delta E$ values for the $N\rightarrow N+1$  electron manifold transitions. Energy differences are plotted by their transition type: $\Delta E^{(0,0)}_{N+1,N}$ (green solid), $\Delta E^{(0,1)}_{N+1,N}$ (green dashed), $ \Delta E^{(1,0)}_{N+1,N}$ (red dashed), $\Delta E^{(1,1)}_{N+1,N}$ (red solid), and the subset of differences involving the second excited state of both manifolds (gray).
        }
    \end{figure}

\begin{figure}[!ht]
        \includegraphics[width=\columnwidth]{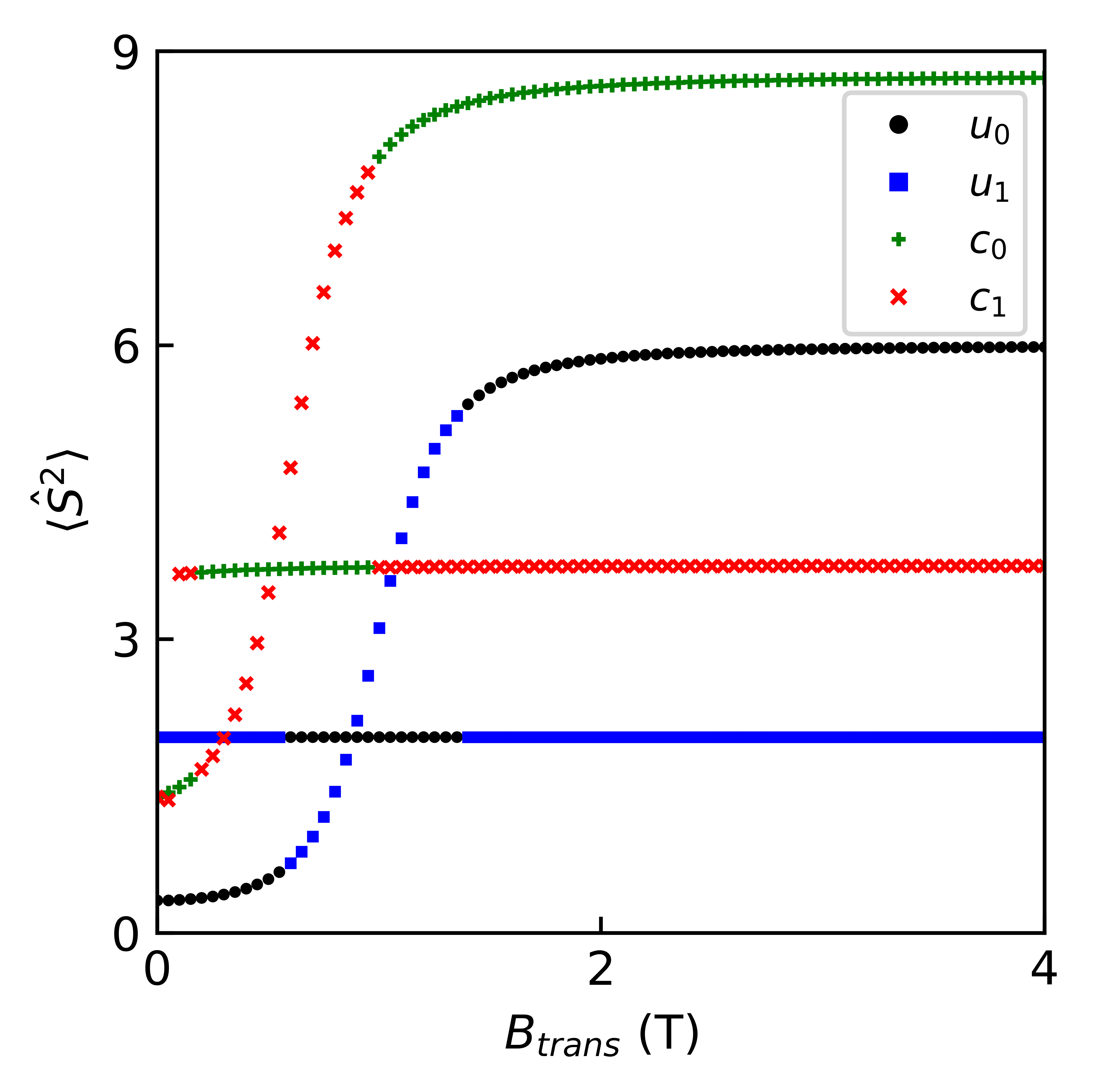}
        \colorcaption{\label{fig:ssquared}
        $S^{2}$ projections of the first two states in the uncharged and charged manifolds.
        }
    \end{figure}

    We further investigate the dependency of the energy levels and energy differences, as a function of the applied transverse field, as shown Fig.~\ref{fig:energetics} and Fig.~\ref{fig:splits}, respectively. 
    Fig.~\ref{fig:energetics} displays interesting level crossing and avoided crossing behavior for applied fields in the range of $B_{trans}=0$ T and $B_{trans}=1.5$ T.
    The interesting level crossing and avoided crossing behavior also appears in the energy differences as shown in Fig.~\ref{fig:splits}. 
    The same level crossings in the range of $B_{trans}=0$ T and $B_{trans}=1.5$ T result in a flip of energetic ordering of the excited state $\Delta E^{(1,1)}_{N+1,N}$ and ground state $\Delta E^{(0,0)}_{N+1,N}$ transitions. 
    Using the eigenstates projected onto the axis corresponding to the magnetic anisotropy Hamiltonian term, the transverse magnetic field mixes states with different total $S^{2}$ and $m_{S}$ spin quantum numbers. The resulting spin eigenvectors are found to primarily have $m_{S}=0$ and a non-trivial $S^{2}$ value as shown in Fig.~\ref{fig:ssquared}.
    
    \subsection{Impact on Differential Conductance}    
\begin{figure}[!ht]
        \includegraphics[width=\columnwidth]{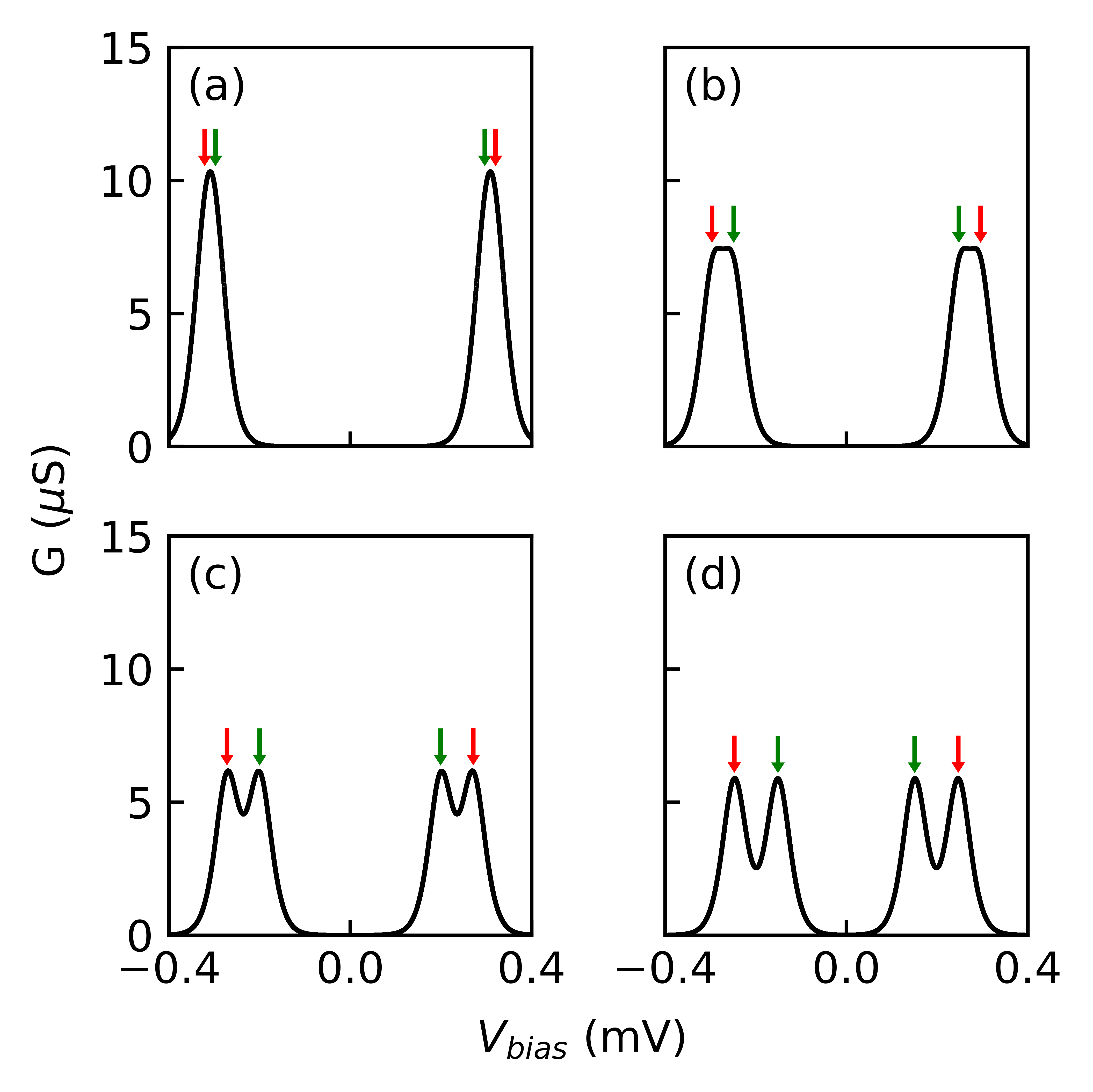}
        \colorcaption{\label{fig:conductance-contributions}
        Differential conductance $G$ as a function of bias voltage for $V_{gate}=0.2$ mV. 
        Four peaks of conductance correspond with key energy difference values entering the bias window, $\pm2\Delta E^{(0,0)}_{N+1,N}$ (green arrow) and $\pm2\Delta E^{(1,1)}_{N+1,N}$ (red arrow), broadened by temperature. 
        The values of $J_{1i}$ (in units of $\text{cm}^{-1}$) are (a) $-0.2$, (b) $-0.4$, (c) $-0.6$, and (d) $-0.8$.
        }
    \end{figure}

    Next we solve the current equation Eq.~(\ref{eqn:electric-current-bylead}) for the system used in the prior section in order to demonstrate the impact of the spin Hamiltonian eigenvalue differences on differential conductance.
    We apply a transverse field of 2.0 T, sweep the bias voltage from -1 mV to +1 mV, and numerically differentiate the current with respect to the bias voltage to obtain predicted differential conductance. 
    The results are shown in Fig.~\ref{fig:conductance-contributions}.
    Changing the spin Hamiltonian parameters results in different conductance spectra.
    Using the magnitude of $J_{1i}$ parameter as an example, we find two conductance peaks when the electron-dimer exchange coupling is turned off, as would be expected for a system within CB conditions. 
    As the absolute magnitude of the exchange coupling is increased, additional peaks appear. 
    Each peak is found to correspond with the energy differences of the ground states $\Delta E^{(0,0)}_{N+1,N}$ and excited states $\Delta E^{(1,1)}_{N+1,N}$ as the energy differences enter the bias window.
    If the magnitude of $J_{1i}$ is increased, the peaks become more aligned with the value $\pm2\Delta E^{(k,k)}_{N+1,N}$ involving the $k$'th energy states.  
    
\begin{figure}[!ht]
        \includegraphics[width=\columnwidth]{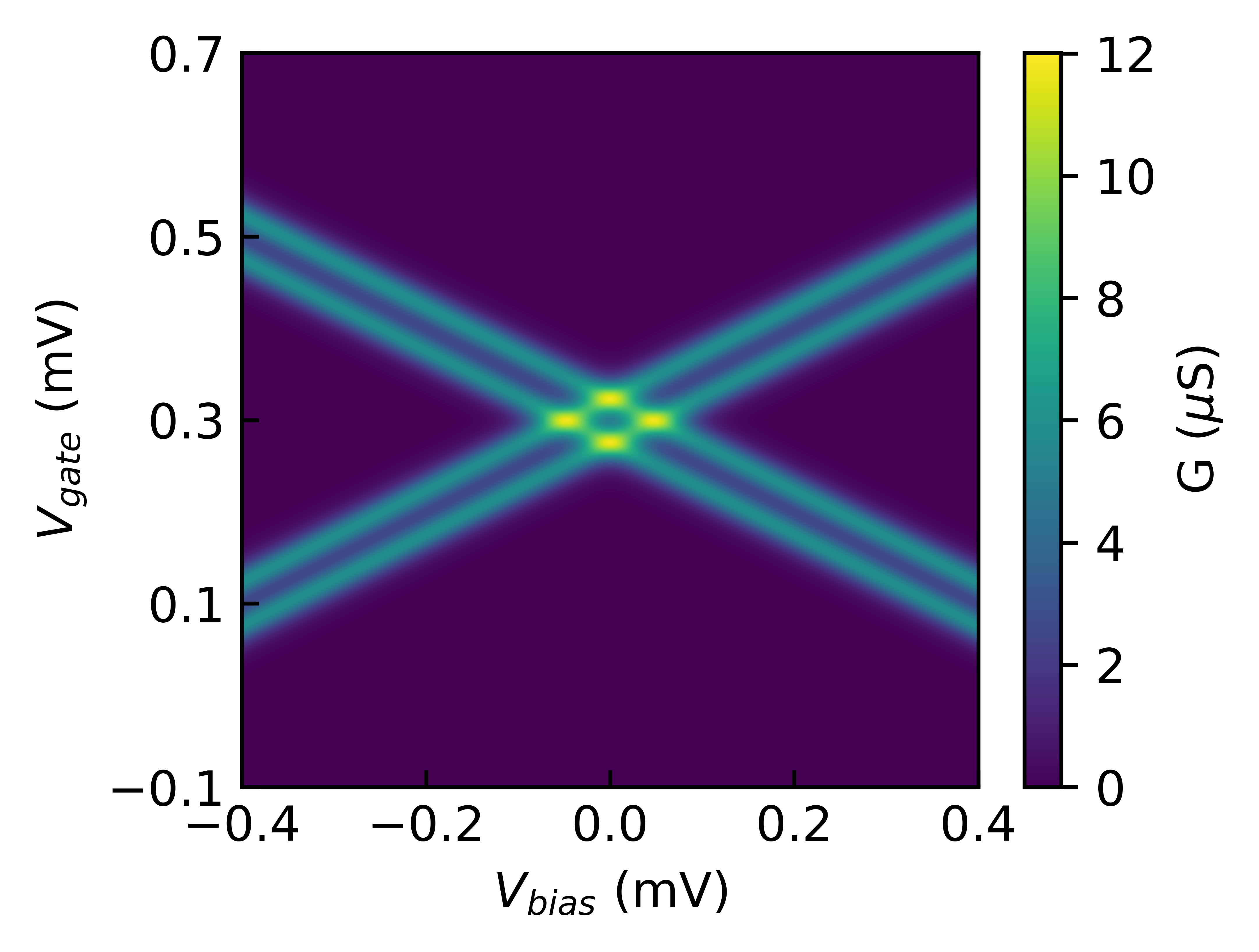}
        \colorcaption{\label{fig:diamond}
        Differential conductance $G$ as a function of bias voltage and gate voltage. 
        For the parameter set chosen, four conductance lines are clearly seen, associated with the inclusion of four energy difference values as bias is increased.
        }
    \end{figure}

    Next we investigate the differential conductance as a function of both gate and bias voltage. The results are shown in Fig.~\ref{fig:diamond}. 
    A Coulomb diamond-like feature appears in the plot, with more than one conductance maxima lines proportional to the magnitude of the bias and gate voltages.
    When the exchange interaction of the itinerant electron with the dimer is turned off and the excitation energies become degenerate, these features disappear, and the typical Coulomb diamond plot is reproduced.

    \subsection{Mapping of Spin Hamiltonian Parameters}
    \label{params}
    The finding from the prior section suggests that one could use differential conductance measurements to characterize and parameterize the spin Hamiltonian.
    The mapping procedure to the model in this work is complicated by four parameter types, $J_{1i}$, $J_{23}$, $D$, and $g_{i}$.
    To make the procedure tractable, we first assume the same simplifications of the last section, but now allow $D$, $J_{23}$, and $J_{1i}$ to take on reasonable values.
    The range of values chosen are provided as an example of the energy range relevant for MMs and QDs. 
    The magnetic anisotropy is allowed to take on a representative ``easy'' axis, no axis, and ``hard'' axis values: $D \in \{-0.6,\;0.0,\;0.6\} \; \text{cm}^{-1}$.
    Similarly, we choose the dimer exchange coupling to be either ferromagnetic or antiferromagnetic: $J_{23} \in \{-0.6,\;0.6\} \; \text{cm}^{-1}$. 
    Last, we choose the dot-dimer exchange coupling to be either ferromagnetic, ``weakly'' ferromagnetic, weakly antiferromagnetic, or antiferromagnetic: $J_{1i} \in \{-0.8,\; -0.08,\;0.08,\;0.8\} \; \text{cm}^{-1}$.
    We assume that the $J_{1i}$ value is non-zero to ensure the QD's coupling to the spin space of the dimer.
    We also assume that the $g$ factor of each spin center in the dimer has the same value, but can be different from the QD's effective $g$ factor of 2.2: $g_{23} \in \{2.2,\;3.2,\;4.2\}$.

\begin{figure}[!ht]
        \includegraphics[width=\columnwidth]{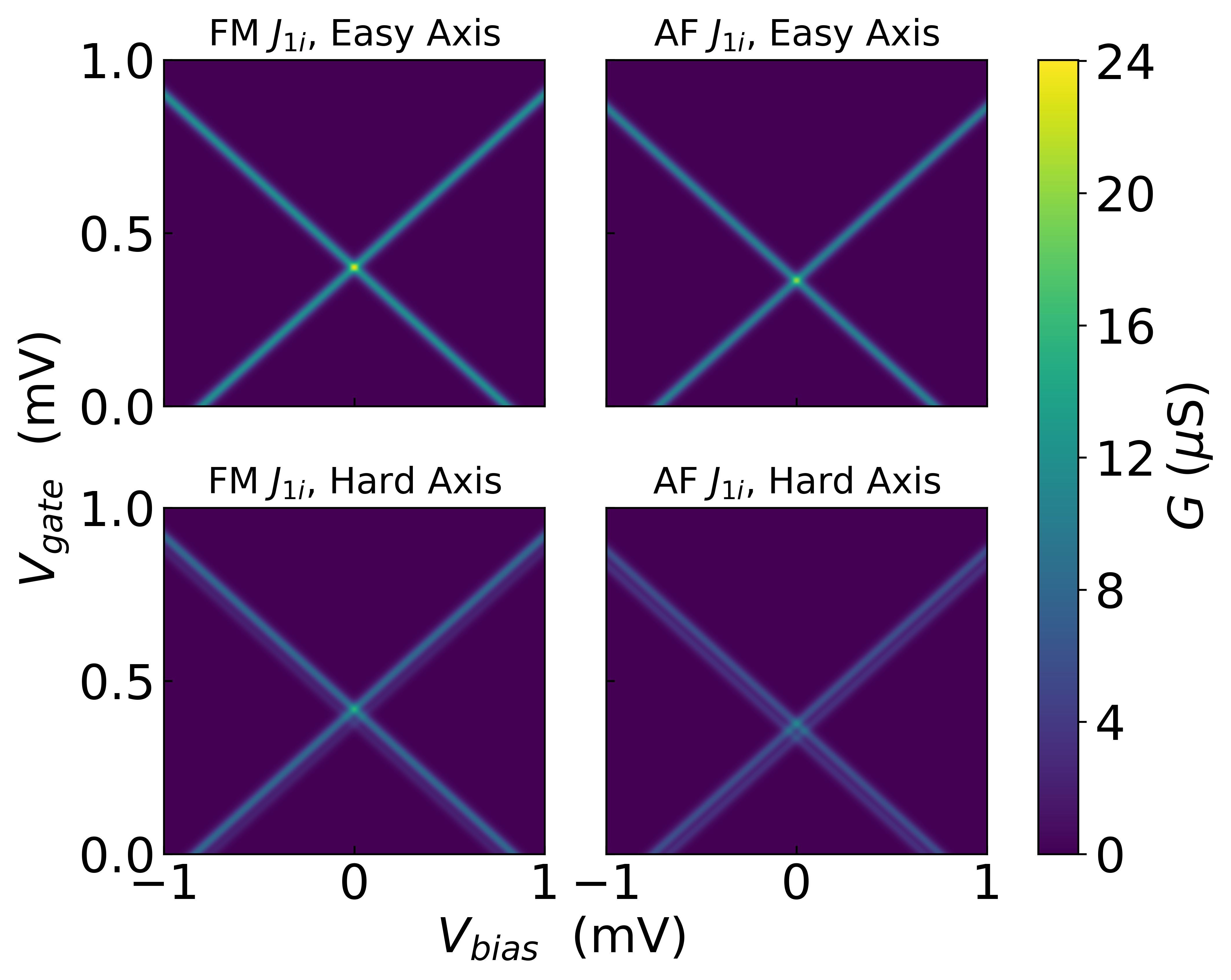}
        \colorcaption{\label{fig:zerofield_examples}
        Differential conductance $G$ as a function of bias voltage and gate voltage, with a ferromagnetic $J_{23}$ parameter. The correspondence of plot labels to numerical parameters can be found in the text of Section \ref{params}.
        }
    \end{figure}

    The first stage in the mapping of the parameters is to make differential conductance measurements without the use of an applied magnetic field. 
    By doing so, the Zeeman term of the Hamiltonian disappears, and the $g$ factor does not need to be parameterized in this stage.
    We explore the parameter space of $J_{1i}$, $J_{23}$, and $D$ using the $V_{gate}$ and $V_{bias}$ independent variables. 
    The key highlights of the dependency of the conductance spectra on the sign of $J_{1i}$ and the sign of $D$ is highlighted in Fig.~\ref{fig:zerofield_examples}.

\begin{figure*}[!ht]
        \includegraphics[height=3in]{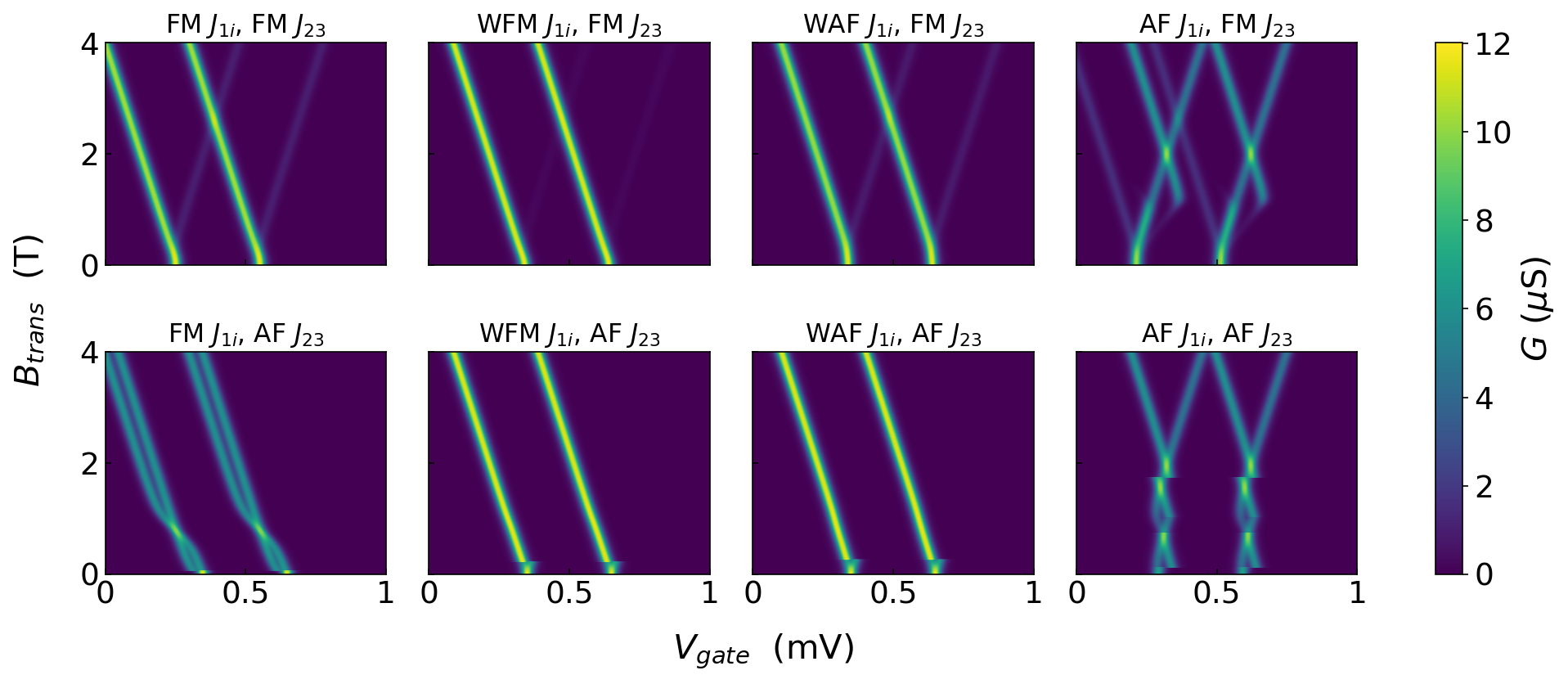}
        \colorcaption{\label{fig:withfield_examples}
        Differential conductance $G$ as a function of gate voltage and transverse applied magnetic field, with an easy axis $D$. The correspondence of plot labels to numerical parameters can be found in the text of Section \ref{params}.
        }
    \end{figure*}

    The next stage in mapping of the parameters is to utilize the Zeeman field to investigate Hamiltonian terms that should be sensitive to the field magnitude.
    We also set $B_{trans} = 2.0$ T for those independent variable combinations that do not involve it, such as the dependence of conductance on $B_{z}$ and $V_{bias}$. 
    We find important changes in the conductance to identify the sign and magnitude of $J_{1i}$ and the sign of $J_{23}$ as shown in Fig.~\ref{fig:withfield_examples}. 
    
\begin{figure*}[!ht]
        \includegraphics[height=2.73in]{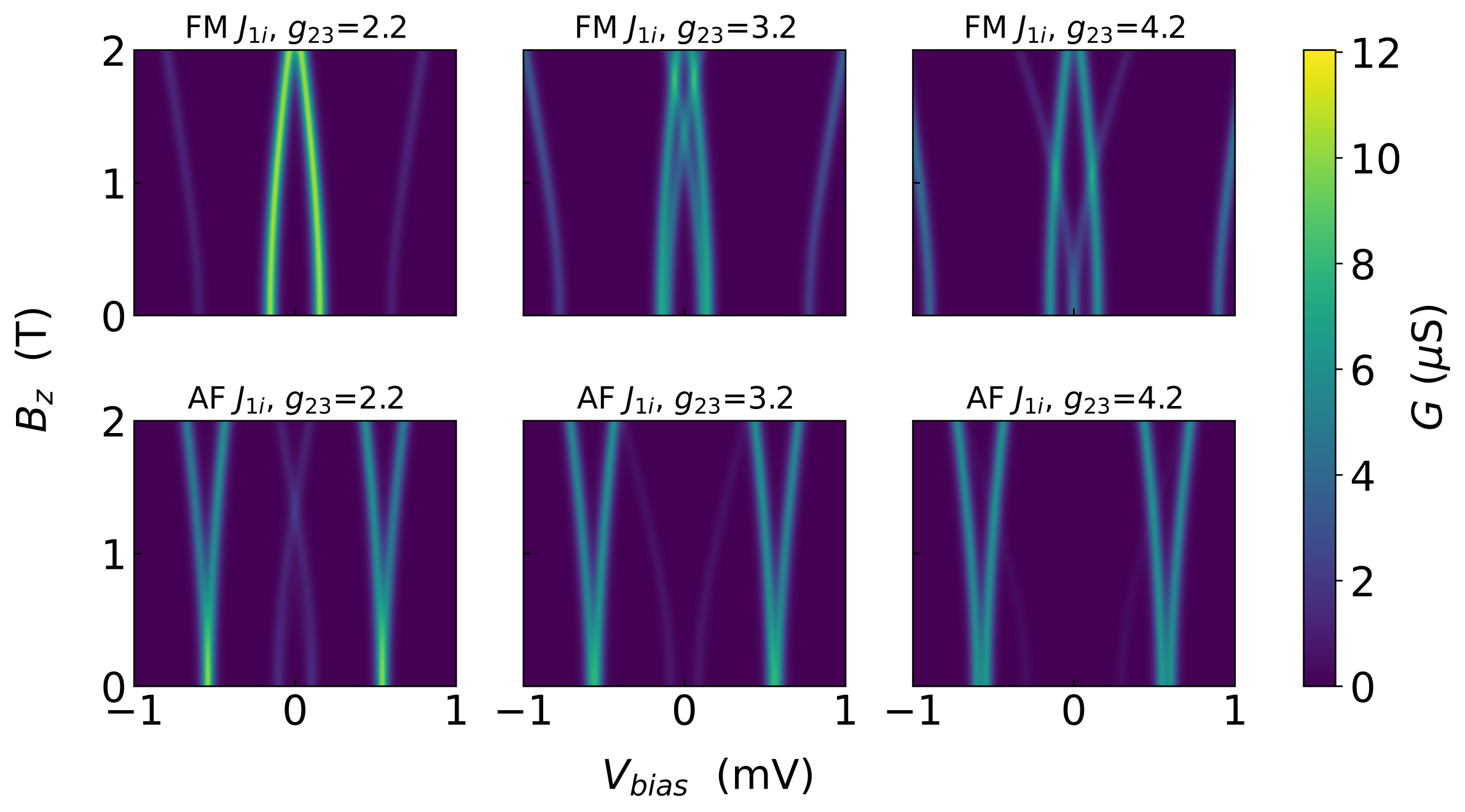}
        \colorcaption{\label{fig:gfactor}
        Differential conductance $G$ as a function of bias voltage and parallel applied magnetic field, with an easy axis $D$ and a ferromagnetic $J_{23}$. The correspondence of plot labels to numerical parameters can be found in the text of Section \ref{params}.
        }
    \end{figure*}

    Once the signs and magnitudes of $J_{1i}$, $J_{23}$, and $D$ have been determined, we create a third stage of measurements to get information on the $g$ factor.
    We find in Fig.~\ref{fig:gfactor} that it is relatively easy to see changes in the conductance spectra at the resolution of our example using the parallel-aligned applied magnetic field and bias voltage because the change in location and magnitude of the peaks is dependent on the magnitude of the $g$ factor. 

\begin{figure}[!ht]
        \includegraphics[width=\columnwidth]{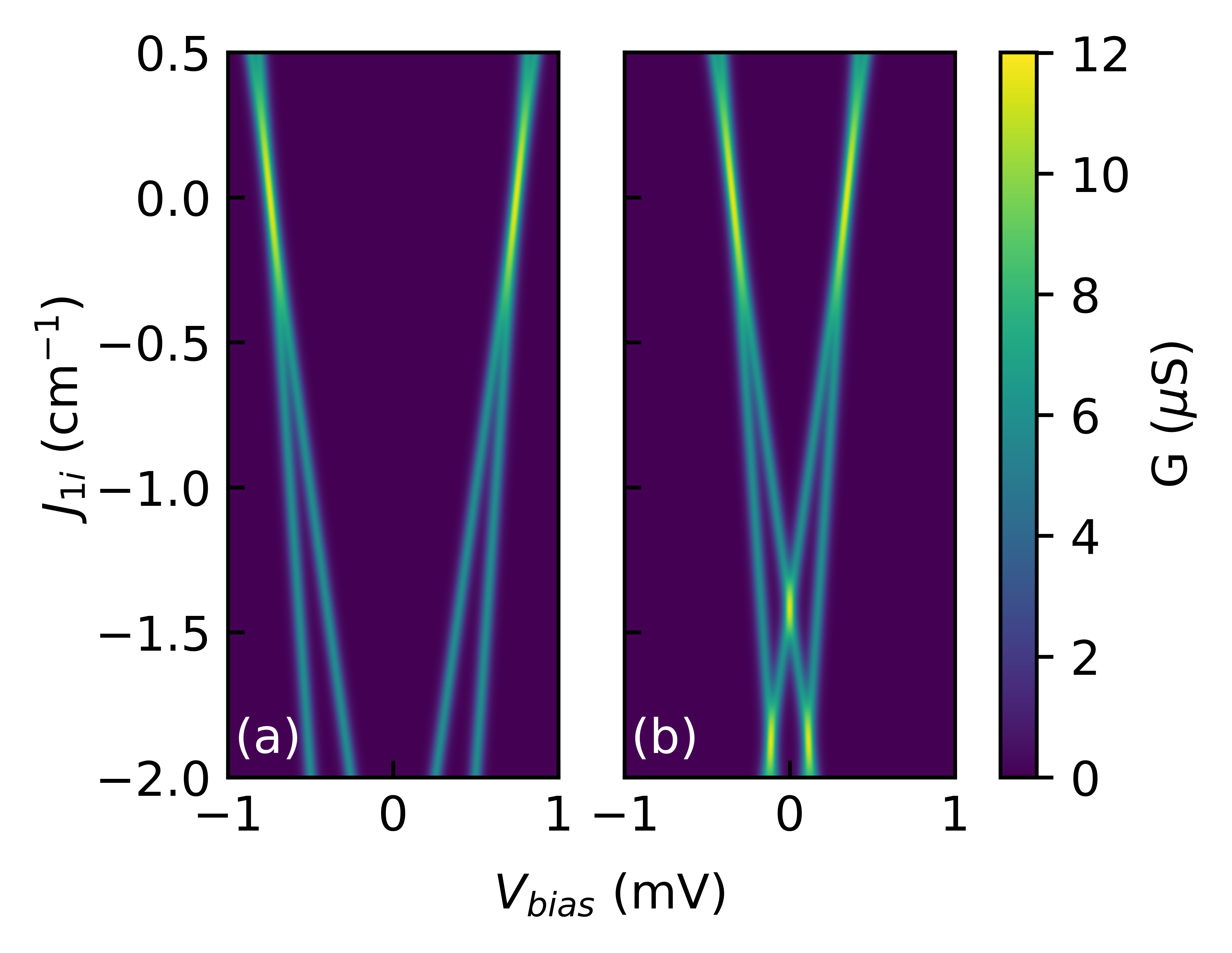}
        \colorcaption{\label{fig:diamond_j1i}
        Differential conductance $G$ as a function of bias voltage and exchange coupling $J_{1i}$. 
        The exchange coupling breaks the symmetry of the conductance peaks as the absolute magnitude is increased from $J_{1i}=0.0\;\text{cm}^{-1}$. 
        Different gate voltage $V_{gate}$ at (a) $0.0$ mV and (b) $0.2$ mV changes the bias needed to split the conductance peaks.
        }
    \end{figure}

    Finally, noting that the sensitivity of the magnitude of the exchange coupling strength of the QD and the spin system is apparent in the prior figures, we further explore the role of the magnitude and sign of $J_{1i}$ in Fig.~\ref{fig:diamond_j1i}. 
    As the absolute magnitude of the exchange coupling is increased from $J_{1i}=0.0\;\text{cm}^{-1}$, the double peak feature is broken into four peaks.
    If a gate voltage is applied, one can access the three-peak regime within a chosen bias window, and even revert back to two conductance peaks for large enough $J_{1i}$.
    The choice of gate voltage is equivalent to accessing different parts of the Coulomb diamond shown in Fig.~\ref{fig:diamond}.

\section{Discussion}
\label{sec:discussion}

    The dependency of the conductance splitting on each spin Hamiltonian parameter is evident in Figures \ref{fig:zerofield_examples}-\ref{fig:diamond_j1i}. The measurement scheme, involving each of the three stages, is the main result of this paper. 
    Parameterizing all four spin Hamiltonian terms at once is a difficult task, and so we find that the three-stage process listed here is a method to constrain the parameter search by simplifying the parameter space at each stage.
    We first note the reason why different peaks and lines appear in the differential conductance plots.
    The splitting of conductance peaks are explained by the inclusion of additional energy differences (and thus additional transport channels) in the bias window, as shown in Fig.~\ref{fig:conductance-contributions}. 
    Both the $\Delta E^{(0,0)}_{N+1,N}$ and $\Delta E^{(1,1)}_{N+1,N}$ energy differences are degenerate at zero exchange coupling, which leads to only two peaks appearing in the differential conductance plots.  
    We have found that this degeneracy appears only when a critical threshold of $B_{trans}$ is reached depending on the parameters of $J_{23}$, $D$, and $g$.

    The features resulting from energy difference degeneracy and change of ordering of the ground states for each charge manifold appear to allow one to probe different regimes for a chosen parameter combination. 
    In Fig.~\ref{fig:zerofield_examples}, we find that for ferromagnetic $J_{23}$, one can determine the sign of $D$ and $J_{1i}$ by comparing the peak magnitude and location of the differential conductance and the number of conductance lines.
    When comparing the lower-left corner of Fig.~\ref{fig:withfield_examples} to Fig.~\ref{fig:ssquared}, we see that the re-ordering of the ground state leads to observed changes in the conductance spectra.
    That particular conductance plot is unique to the parameter choices we explore, in comparison to the other combinations of ferromagnetic and antiferromagnetic $J_{1i}$ and $J_{23}$ for an easy $D$ axis.
    There are still challenges, however, when searching for particular parameter combinations in each stage. 
    For example, if $J_{23}$ is antiferromagnetic, it is difficult to discern differences in the conductance spectra within the first stage with zero applied magnetic field (see Supplementary Fig. 2~\cite{[{See Supplemental Material at }]supp}).

    While an analytical solution for the eigenenergies is not trivial, Fig.~\ref{fig:diamond_j1i} demonstrates that one can still infer important information regarding the contribution of each parameter in the system to the additional conductance peaks.
    It's important to note the role of temperature. For example, with the small $J_{1i}$ in \ref{fig:conductance-contributions}, the peaks of conductance do not align with the energy difference values because of the finite temperature broadening induced by the Fermi functions in Eq.~(\ref{eqn:trans-rate-absorb}) and Eq.~(\ref{eqn:trans-rate-emit}).
    Because of the temperature dependence in the transition rates equation, higher temperatures smooth out the energy differences, and thus the four peak features may be difficult to resolve in a conductance measurement.
    High enough temperature, however, will result in moving out of the regime that utilizes the CB, in which our results are no longer applicable.
    
    We also compare the results of this work to a similar approach that was used for systems that contained a few of the elements of our model Hamiltonian, but not its entirety. 
    In the double dot model of Ref.~\cite{zhang12}, conductance peak splitting was found as a function of the interdot coupling.
    Comparing to our work, a similar type of coupling is accomplished through the $\mathcal{H}_{eS}$ Hamiltonian term. 
    Instead of transport between the two dots in their work, our model has an effective spin interaction mediated between the two centers via $\mathcal{H}_{23}$ and onsite zero-field splitting terms $\mathcal{H}_{A}$, which establishes an energetic preference of states, and thus ordering of preferable transport channels.
    
    An interesting feature can be seen in the energy differences shown in Fig.~\ref{fig:splits}. 
    Sweeping the transverse magnetic field within the parameter space changes the number of state transitions within a given bias window.
    If one were to extend our two electron manifold and eigenstate model to include the new transport channels, additional features in the electronic current should appear.
    Based on the results of Fig.~\ref{fig:conductance-contributions}, as long as the lowest transition $\Delta E^{(i,0)}_{N+1,N}$ for $i > 2$ is within the bias window, these additional state transitions would result in a change of the width and number of the predicted conductance peaks.
     
    Our model can be extended to include more effects found in systems consisting of a QD coupled to a molecular complex or impurity. 
    For example, one could introduce the effects of charging on the coupled spin system. 
    It's known that the SMM $\text{Mn}_{12}$ zero-field splitting parameter and other spin parameters change upon charging \cite{basler05}.
    We find that accommodating for a charge state anisotropy in zero-field splitting results in similar or different conductance peak behavior depending on the parameter range.
    If we use the same parameters in Fig.~\ref{fig:diamond}, and instead choose an uncharged $D_{0}=-0.6\;\text{cm}^{-1}$ and charged $D_{1} \gtrapprox -1.2\;\text{cm}^{-1}$, the four peak feature is retained. 
    If one instead chooses $D_{1} \lessapprox -1.2\;\text{cm}^{-1}$, the four peak feature disappears.
    
\section{Summary and Outlook}
\label{sec:summary}
    As has been shown, within the CB for the model of an itinerant electron originating from leads passing into a QD connected to a multi-spin complex, conductance peak splitting appears within a range of parameters related to the magnetic response of the central region.
    We have also shown that one may match the location and number of differential conductance peaks as a function of bias voltage, gate voltage, and anisotropically-applied magnetic field in order to effectively measure each parameter of the model spin Hamiltonian.
    This provides a mapping to experimentally determine these magnetic properties for increasingly small nanoscale devices, such as molecular transistors, using only electronic differential conductance.
    
    One challenge with this mapping is the number of Hamiltonian parameters.
    Machine-learning methods may be used to help fit experimental differential conductance measurements to the model Hamiltonian explored in this work, using the results from this work as the starting point for a training set. 
    In addition, sources of decoherence, the role of the magnitude of decoherence, and the impacts of higher spin $S_{i}>1$ were not studied in this work. 
    Investigations are needed to determine the relative impact of each property on the measured conductance and location of conductance peak splitting. 
    Furthermore, this work assumes a steady-state transport measurement with unpolarized leads. 
    From prior studies \cite{switzer21,switzer22}, one can hypothesize that the itinerant electron will entangle the spin system and produce a time-dependent coherent current in short timescales.
    This time-dependent entanglement may be useful to quantum information science applications.
    Nano- and femto-timescale electron and spin current studies with polarized (and possibly asymmetric) leads are needed to fully explore that possibility.

\begin{acknowledgments}
    We thank Duy Le and Dave Austin for helpful discussions. This work was supported by the Center for Molecular Magnetic Quantum Materials, an Energy Frontier Research Center funded by the U.S. Department of Energy, Office of Science, Basic Energy Sciences under Award No. DE-SC0019330. The authors declare no competing financial interests.
\end{acknowledgments}


\begin{appendix}
\section{Derivation of Density Matrix Elements}
\label{appendix:density-matrix}

    To obtain the density matrix equations and transition rates of our model, we express the coupling terms of $V$ as products of the lead and central region operators. We designate an index notation that tracks all combinations of different $\hat{c}$ and $\hat{d}$ operators,
    \begin{align}
        \label{eqn:perturb-deconstucted}
        V = \sum_{i}t_{i}\hat{F}_{i}\hat{Q}_{i},
    \end{align}
    where $\hat{F}_{i}$ refers to a possible form of the $\hat{c}$ operator, $\hat{Q}_{i}$ refers to a possible form of the $\hat{d}$ operator, and $t_{i}$ is the coupling constant for that combination. 
    
    In this form, the correlators are defined \cite{blum12},
    \begin{align}
        \nonumber
        \Gamma^{+}_{mkln}&=-\frac{1}{\hbar^{2}}\sum_{ij}t_{i}t_{j}\mel{m}{Q_{i}}{k}\mel{l}{Q_{j}}{n}\\
        \label{eqn:plus-correlator}
        &\;\;\;\;\;\;\;\;\times\int_{0}^{\infty} \dd{t}e^{-i\omega_{ln}t}\expval{F_{i}(t)F_{j}},\\
        \nonumber
        \Gamma^{-}_{mkln}&=-\frac{1}{\hbar^{2}}\sum_{ij}t_{i}t_{j}\mel{m}{Q_{j}}{k}\mel{l}{Q_{i}}{n}\\
        \label{eqn:minus-correlator}
        &\;\;\;\;\;\;\;\;\times\int_{0}^{\infty} \dd{t}e^{-i\omega_{mk}t}\expval{F_{j}F_{i}(t)},
    \end{align}
    with the $Q$ operators acting on the Fock spin space of the central region, on the system eigenstates $m$, $k$, $l$, and $n$ with eigenenergies $E_{m}$, $E_{k}$, $E_{l}$, and $E_{n}$,  respectively.
    The leading contribution of the transition rates from central region eigenstate $n$ to $m$ (corresponding to sequential tunneling) is then,
    \begin{align}
        \nonumber
        W_{n'n} &= \Gamma^{+}_{nn'n'n}+\Gamma^{-}_{nn'n'n} \\
        \nonumber
        &= \frac{2\pi}{\hbar}\sum_{iNN'}\left|\mel{n'N'}{t_{i}\hat{F}_{i}\hat{Q}_{i}}{nN}\right|^{2}\mel{N}{\rho_{\text{leads}}(0)}{N}\\
        \label{eqn:transition-rate}
        &\;\;\;\;\;\;\;\;\;\;\;\;\;\;\;\;\times\delta\left(E_{N}-E_{N'}-\hbar\omega_{n'n}\right),
    \end{align}
    where $\omega_{n'n}\equiv(E_{n'}-E_{n})/\hbar$, and the system eigenstates have been expanded in terms of the combined lead $N'$, $N$ and central region $n'$, $n$ eigenstate indices.
    Because of the form of $V$, the only non-zero $W_{n'n}$ elements are those from a charged to an uncharged state or vice-versa (i.e., $W_{u_{i}u_{j}}=W_{c_{i}c_{j}}=0$ $\forall$ $i, j$).
    The damping factor has the form,
    \begin{align}
        \label{eqn:gamma-before-simple}
        \gamma_{n'n}&=\sum_{m}\left[\Gamma^{+}_{n'mmn'}+\Gamma^{-}_{nmmn}\right]-(\Gamma^{+}_{nnn'n'}+\Gamma^{-}_{nnn'n'}).
    \end{align}
    This is simplified by redefining $\gamma_{mm'}$ as is done in Ref.~\cite{gonzalez07} to the form of Eq.~(\ref{eqn:gamma-with-T2}) in the main text.
    
    Finally, clarifying the diagonal versus the off-diagonal terms of the system density matrix, one obtains,
    \begin{align}
        \nonumber
        \dot{\rho}_{nn}(t) &= \frac{i}{\hbar}\left[\rho(t),\mathcal{H}_{0}\right]_{nn}+\sum_{m,n\neq m}\rho_{mm}(t)W_{nm}\\
        \label{eqn:rho-nn}
        &\;\;\;\;-\rho_{nn}(t)\sum_{m,n\neq m}W_{mn},\\
        \label{eqn:rho-nprimen}
        \dot{\rho}_{n'n}(t) &= \frac{i}{\hbar}\left[\rho(t),\mathcal{H}_{0}\right]_{n'n}-\gamma_{n'n}\rho_{n'n}(t).
    \end{align}
    The equation for the dynamics of $\rho(t)$ are given by the Pauli master equation for the diagonal elements, while the off-diagonal elements contain the decoherence of the system with the surrounding reservoir.

    As mentioned in the main text, in order to produce relevant predictions from the generalized master equation, we look at a time range in which the overall relaxation time due to transitions is much longer than the decay of the off-diagonal elements. Because of the time range concerned, $\lim_{t\rightarrow \tau}\dot{\rho}_{n'n}(t)=0$, and so,
    \begin{align}
        \label{eqn:rho-simplified}
        \rho_{n'n}(t)=\frac{i}{\hbar\gamma_{n'n}}\left[\rho(t),\mathcal{H}_{0}\right]_{n'n}.
    \end{align}
    Substituting the central region and leads Hamiltonian into Eq.~(\ref{eqn:rho-nprimen}), one obtains closed equations for the off-diagonal density matrix elements. 
    The off-diagonal terms in the same charge sector, i.e., $n,n'\in\{u_{0},u_{1}\}$ or $n,n'\in\{c_{0},c_{1}\}$, are,
    \begin{align}
        \label{eqn:rho-nprimen-simplified}
        \rho_{n'n}(t)=
        \frac{\mathcal{H}_{n'n}\left(\rho_{n'n'}(t)-\rho_{nn}(t)\right)}{\mathcal{H}_{n'n'}-\mathcal{H}_{nn}-i\hbar\gamma_{n'n}}.
    \end{align}
    Inserting this result into the generalized master equation's diagonal elements, and noting that $\gamma_{nn'}=\gamma_{n'n}$, results in,
    \begin{align}
        \nonumber
        \dot{\rho}_{n'n'}(t)&=\Gamma_{n'n}\left(\rho_{nn}(t)-\rho_{n'n'}(t)\right)\\
        \label{eqn:rho-nprimenprime-simplified}
        &\;\;\;\;\;\;\;\;+\sum_{m,n'\neq m}\rho_{mm}W_{n'm}-\rho_{n'n'}\sum_{m,n'\neq m}W_{mn'},
    \end{align}
    where the Lorentzian decoherence factor is defined as,
    \begin{align}
        \label{eqn:decoherence-factor}
        \Gamma_{n'n} &= \frac{\abs{\mathcal{H}_{n'n}}^{2}}{\hbar^{2}}\frac{2\gamma_{n'n}}{\left(\mathcal{H}_{n'n'}-\mathcal{H}_{nn}\right)^{2}/\hbar^{2}+\gamma_{n'n}^{2}}.
    \end{align}
    The quantity $\gamma_{n'n}^{2}$ can now be interpreted as the broadening factor of the Lorentzian, and the peak of the Lorentzian is maximized if $\mathcal{H}_{n'n'}=\mathcal{H}_{nn}$.
    We point out that while the $n'n$ elements in Eq.~(\ref{eqn:rho-nprimenprime-simplified}) are constrained to the same charge sector, the sums over the index $m$ include only those terms that connect different charge sectors, with the form $W_{u_{i}c_{j}}$ and $W_{c_{j}u_{i}}$.
    This means that the transition rates between any of the different levels between different charge states should be accounted for, if not forbidden by transition rules (e.g., through spin conservation).
    The transition rates from an uncharged eigenstate $u_{j}$ to charged eigenstate $c_{i}$ is derived to be,
    \begin{align}
        \label{eqn:transition-rates}
        W_{c_{i}u_{j}} &= \sum_{\alpha\sigma}W^{\alpha\sigma}_{c_{i}u_{j}},
    \end{align}
    where,
    \begin{align}
        \nonumber
        W^{\alpha\sigma}_{c_{i}u_{j}} &= \frac{2\pi}{\hbar}\nu_{\alpha\sigma}\abs{t_{\alpha\sigma}}^{2}\abs{\mel{c_{i}}{\hat{c}^{\dagger}_{\alpha \sigma}}{u_{j}}}^{2}\\
        &\;\;\;\;\times\int\dd{E}D(E)f_{\alpha}(\Delta E^{(i,j)}_{N+1,N}+E).
    \end{align}
    In the rate equation, $D(E)\equiv D_{\alpha\sigma}(E)D_{\alpha\sigma}(\Delta_{ij}+E)$, and the zero of the chemical potential is measured against the zero of the charged sector. The reverse process has a similar form, 
    \begin{align}
        \nonumber
        W^{\alpha\sigma}_{u_{i}c_{j}} &= \frac{2\pi}{\hbar}\nu_{\alpha\sigma}\abs{t_{\alpha\sigma}}^{2}\abs{\mel{u_{i}}{\hat{c}_{\alpha \sigma}}{c_{j}}}^{2}\\
        &\;\;\;\;\times\int\dd{E}D(E)\left(1-f_{\alpha}(\Delta E^{(j,i)}_{N+1,N}+E)\right).
    \end{align}
    These rates are further simplified by assuming that transport primarily occurs with electrons near the Fermi level of the leads, and so we assign the tunneling rate $w_{\alpha\sigma}=2\pi\abs{t_{\alpha\sigma}}^{2}D(E_{f})/\hbar$.
    Inserting these results along with the steady-state case assumption allows us to obtain closed equations of the density matrix elements.
    \onecolumngrid
    The derived density matrix elements using Eq.~(\ref{eqn:rho-nprimenprime-simplified}) are,
    \begin{align}
        \nonumber
        \eta\rho_{u_{0}u_{0}}&=W_{c_{0}u_{1}} \left(W_{u_{0}c_{0}} \left(\Gamma_{c_{0}c_{1}}+W_{u_{0}c_{1}}+W_{u_{1}c_{1}}\right)+\Gamma_{c_{0}c_{1}} W_{u_{0}c_{1}}\right)\\
        \nonumber&+W_{c_{1}u_{1}} \left(\Gamma_{c_{0}c_{1}} \left(W_{u_{0}c_{0}}+W_{u_{0}c_{1}}\right)+W_{u_{0}c_{1}} \left(W_{u_{0}c_{0}}+W_{u_{1}c_{0}}\right)\right)\\
        &+\Gamma_{u_{0}u_{1}} \left(\Gamma_{c_{0}c_{1}} \left(W_{u_{0}c_{0}}+W_{u_{0}c_{1}}+W_{u_{1}c_{0}}+W_{u_{1}c_{1}}\right)+\left(W_{u_{0}c_{0}}+W_{u_{1}c_{0}}\right) \left(W_{u_{0}c_{1}}+W_{u_{1}c_{1}}\right)\right),
    \end{align}
    \begin{align}
        \nonumber
        \eta\rho_{u_{1}u_{1}}&=W_{c_{0}u_{0}} \left(\Gamma_{c_{0}c_{1}} \left(W_{u_{1}c_{0}}+W_{u_{1}c_{1}}\right)+W_{u_{1}c_{0}} \left(W_{u_{0}c_{1}}+W_{u_{1}c_{1}}\right)\right)\\
        \nonumber&+W_{c_{1}u_{0}} \left(\Gamma_{c_{0}c_{1}} \left(W_{u_{1}c_{0}}+W_{u_{1}c_{1}}\right)+W_{u_{1}c_{1}} \left(W_{u_{0}c_{0}}+W_{u_{1}c_{0}}\right)\right)\\
        &+\Gamma_{u_{0}u_{1}} \left(\Gamma_{c_{0}c_{1}} \left(W_{u_{0}c_{0}}+W_{u_{0}c_{1}}+W_{u_{1}c_{0}}+W_{u_{1}c_{1}}\right)+\left(W_{u_{0}c_{0}}+W_{u_{1}c_{0}}\right) \left(W_{u_{0}c_{1}}+W_{u_{1}c_{1}}\right)\right),
    \end{align}
    \begin{align}
        \nonumber
        \eta\rho_{c_{0}c_{0}}&=W_{c_{0}u_{0}} \left(W_{c_{0}u_{1}} \left(\Gamma_{c_{0}c_{1}}+W_{u_{0}c_{1}}+W_{u_{1}c_{1}}\right)+W_{c_{1}u_{1}} \left(\Gamma_{c_{0}c_{1}}+W_{u_{0}c_{1}}\right)+\Gamma_{u_{0}u_{1}} \left(\Gamma_{c_{0}c_{1}}+W_{u_{0}c_{1}}+W_{u_{1}c_{1}}\right)\right)\\
        \nonumber&+W_{c_{0}u_{1}} \left(\left(\Gamma_{c_{0}c_{1}}+W_{u_{1}c_{1}}\right) \left(W_{c_{1}u_{0}}+\Gamma_{u_{0}u_{1}}\right)+W_{u_{0}c_{1}} \Gamma_{u_{0}u_{1}}\right)\\
        &+\Gamma_{c_{0}c_{1}} \left(\Gamma_{u_{0}u_{1}} \left(W_{c_{1}u_{0}}+W_{c_{1}u_{1}}\right)+W_{c_{1}u_{0}} W_{c_{1}u_{1}}\right),
    \end{align}
    and,
    \begin{align}
        \nonumber
        \eta\rho_{c_{1}c_{1}}&=W_{c_{0}u_{1}} \left(\Gamma_{c_{0}c_{1}} \left(W_{c_{0}u_{0}}+\Gamma_{u_{0}u_{1}}\right)+W_{c_{1}u_{0}} \left(\Gamma_{c_{0}c_{1}}+W_{u_{0}c_{0}}\right)\right)\\
        \nonumber&+\Gamma_{u_{0}u_{1}} \left(\Gamma_{c_{0}c_{1}} W_{c_{0}u_{0}}+\left(W_{c_{1}u_{0}}+W_{c_{1}u_{1}}\right) \left(\Gamma_{c_{0}c_{1}}+W_{u_{0}c_{0}}+W_{u_{1}c_{0}}\right)\right)\\
        &+W_{c_{1}u_{1}} \left(W_{c_{0}u_{0}} \left(\Gamma_{c_{0}c_{1}}+W_{u_{1}c_{0}}\right)+W_{c_{1}u_{0}} \left(\Gamma_{c_{0}c_{1}}+W_{u_{0}c_{0}}+W_{u_{1}c_{0}}\right)\right),
    \end{align}
    where $\eta$ is a normalization constant.
    \twocolumngrid
\end{appendix}

\bibliography{biblio}

\end{document}



\title{Supplementary to Conductance Peak Splitting in the Coulomb Blockade as Signature of Spin Interactions}

\author{Eric D. Switzer}
\affiliation{Department of Physics, University of Central Florida, Orlando, Florida 32816, USA}
\author{Xiao-Guang Zhang}
\affiliation{Department of Physics, Center for Molecular Magnetic Quantum Materials and Quantum Theory Project, University of Florida, Gainesville, Florida 32611, USA}
\author{Volodymyr Turkowski}
\affiliation{Department of Physics, University of Central Florida, Orlando, Florida 32816, USA}
\author{Talat S. Rahman}
\email[Corresponding author email: ]{talat.rahman@ucf.edu}
\affiliation{Department of Physics, University of Central Florida, Orlando, Florida 32816, USA}

\date{\today}

\maketitle
\onecolumngrid

\section{Plots of Conductance For No Applied Magnetic Field}

    \begin{figure}[!ht]
        \includegraphics[width=\columnwidth]{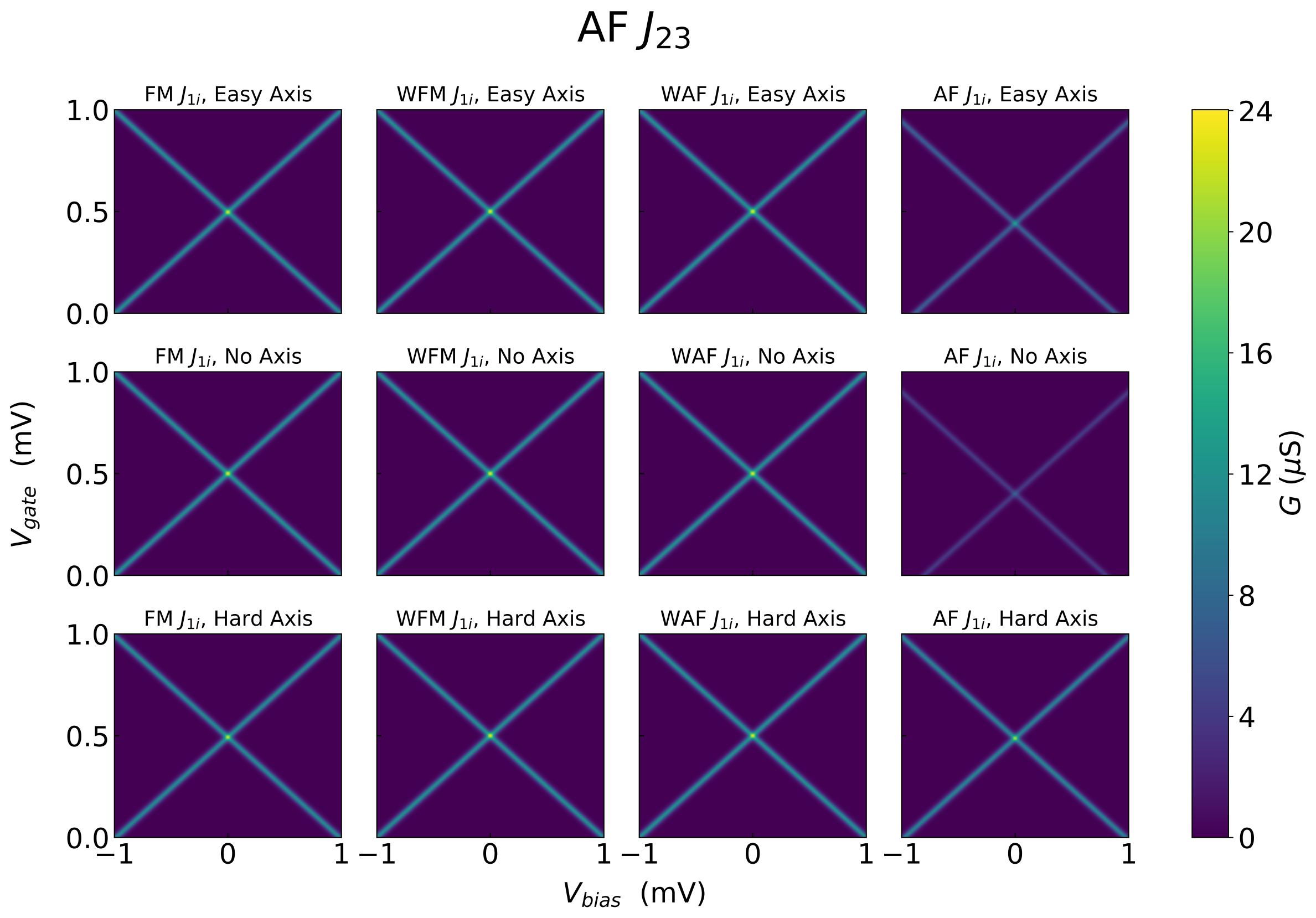}
        \colorcaption{
        Differential conductance $G$ as a function of bias voltage and gate voltage for antiferromagnetic $J_{23}$.
        The correspondence of plot labels to numerical parameters can be found in the main text.
        }
    \end{figure}

    \begin{figure}[!ht]
        \includegraphics[width=\columnwidth]{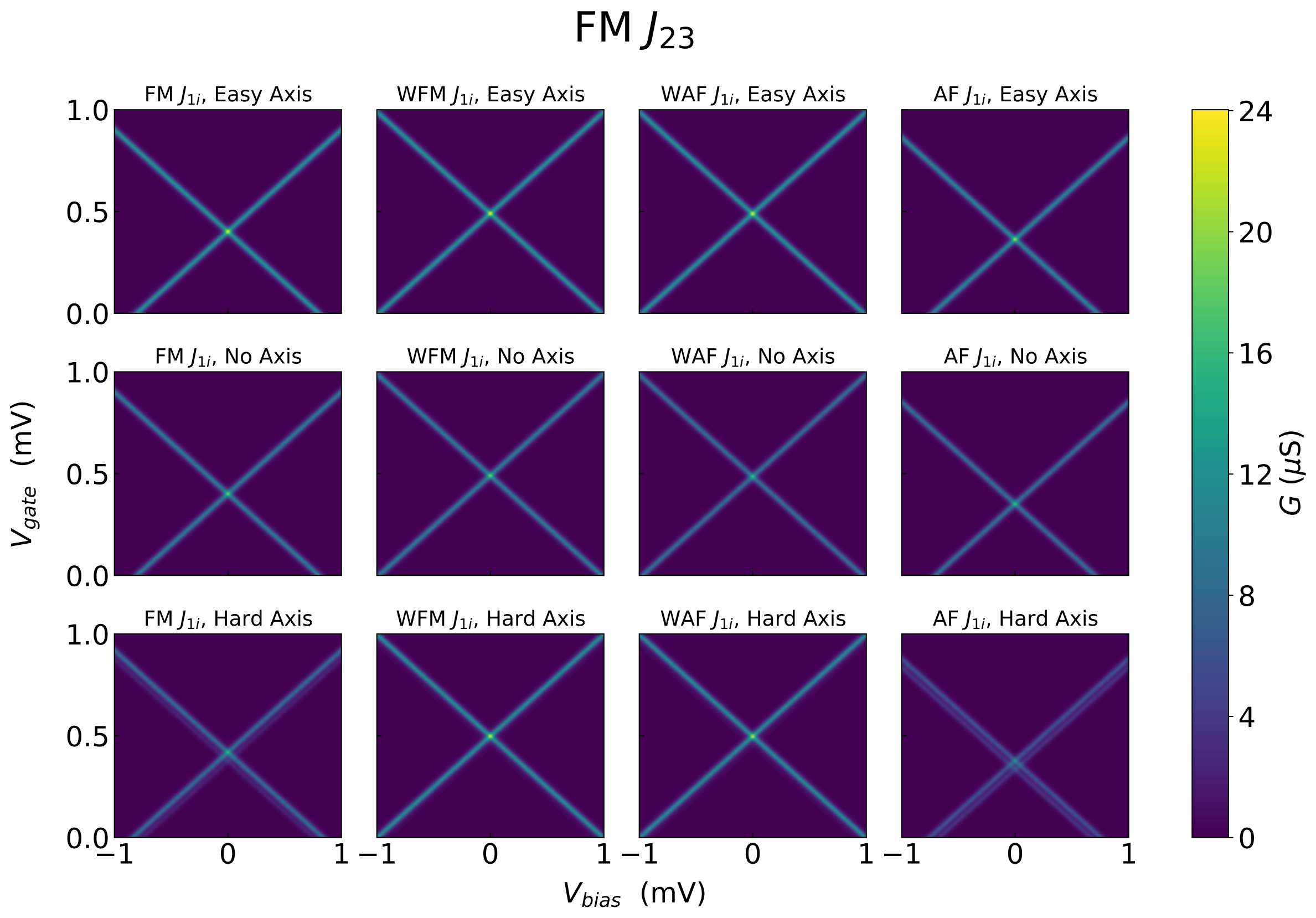}
        \colorcaption{
        Differential conductance $G$ as a function of bias voltage and gate voltage for ferromagnetic $J_{23}$.
        The correspondence of plot labels to numerical parameters can be found in the main text.
        }
    \end{figure}

\section{Plots Conductance With an Applied Magnetic Field}

    \begin{figure}[!ht]
        \includegraphics[width=\columnwidth]{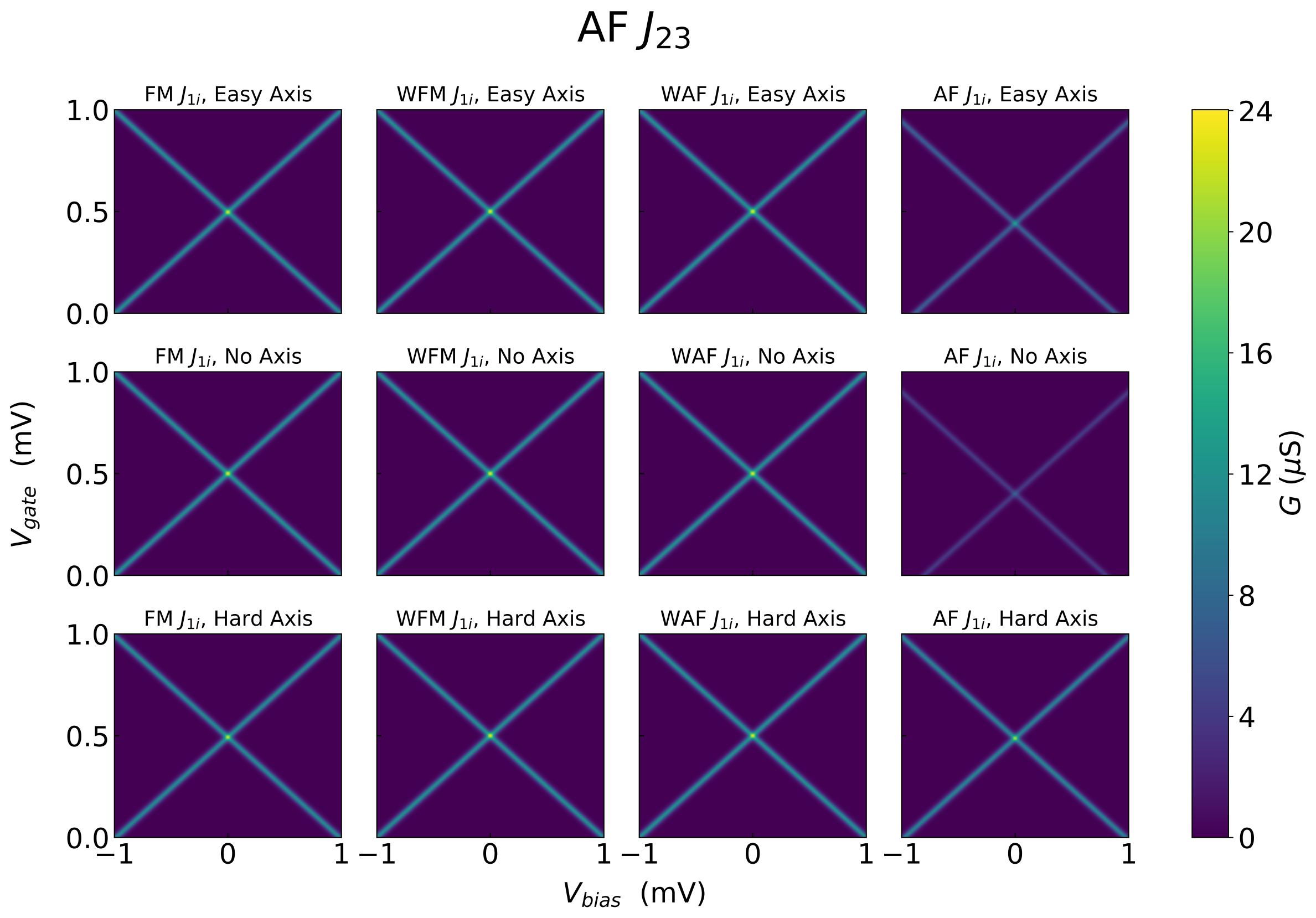}
        \colorcaption{
        Differential conductance $G$ as a function of gate voltage and transverse applied magnetic field for ferromagnetic $J_{23}$.
        The correspondence of plot labels to numerical parameters can be found in the main text.
        }
    \end{figure}

    \begin{figure}[!ht]
        \includegraphics[width=\columnwidth]{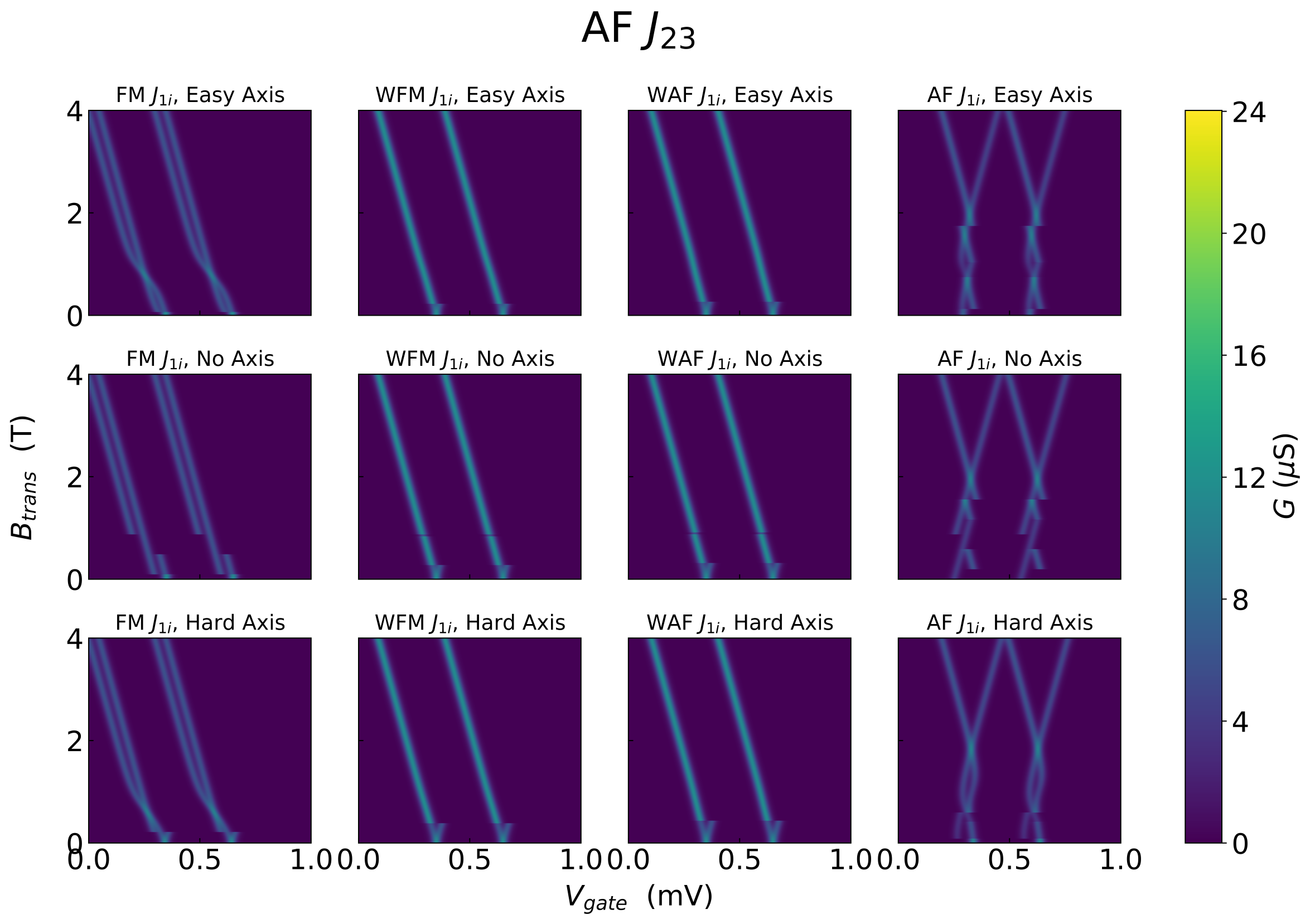}
        \colorcaption{
        Differential conductance $G$ as a function of gate voltage and transverse applied magnetic field for antiferromagnetic $J_{23}$.
        The correspondence of plot labels to numerical parameters can be found in the main text.
        }
    \end{figure}